\documentclass{article}

\usepackage{arxiv}

\usepackage{amsmath}
\usepackage{graphicx}
\usepackage{dcolumn}
\usepackage{bm}
\usepackage{amssymb}

\usepackage[utf8]{inputenc}
\usepackage[T1]{fontenc}
\usepackage{mathptmx}
\usepackage{etoolbox}
\usepackage{caption}
\usepackage{subcaption}

\usepackage[T1]{fontenc}    
\usepackage{hyperref}       
\usepackage{nameref}        
\usepackage{url}            
\usepackage{booktabs}       
\usepackage{amsfonts}       
\usepackage{nicefrac}       
\usepackage{microtype}      
\usepackage{multirow}
\usepackage{fancyhdr}       
\usepackage{graphicx}    
\usepackage{geometry}
\usepackage{xcolor}
\usepackage{algorithm}
\usepackage{algorithmic}

\usepackage{float}

\usepackage{booktabs}
\usepackage{enumitem}
\usepackage{lineno}

\usepackage{nomencl}
\makenomenclature
\usepackage{etoolbox}

\renewcommand\nomgroup[1]{%
  \item[\bfseries
  \ifstrequal{#1}{M}{Structural Parameters}{%
  \ifstrequal{#1}{G}{Geometry}{%
  \ifstrequal{#1}{P}{Process Parameters}{%
  \ifstrequal{#1}{V}{Velocities \& Friction}{%
  \ifstrequal{#1}{D}{Dimensional}{%
  \ifstrequal{#1}{N}{Non-dimensional}{%
  \ifstrequal{#1}{C}{Continuation}{%
  \ifstrequal{#1}{S}{Stochastic Parameters}{%
  \ifstrequal{#1}{E}{Entropy}{%
  \ifstrequal{#1}{A}{Acronyms}{%
  \ifstrequal{#1}{B}{Basin Stability Analysis}{%
  \ifstrequal{#1}{L}{Linear Stability Analysis}{%
  \ifstrequal{#1}{F}{Forces}{}}}}}}}}}}}}}%
]}
\usepackage[numbers,sort&compress]{natbib}
\usepackage{doi}
\usepackage{multicol}
\usepackage{authblk}
\usepackage{hyperref}

\title{Interplay of Nonsmoothness, Time Delay, and Stochasticity in Turning Dynamics}

\date{}

\newbox{\orcid}\sbox{\orcid}{\includegraphics[scale=0.06]{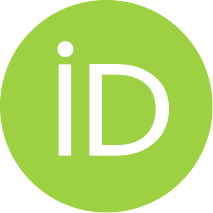}}

\author[1]{%
	\href{https://orcid.org/0000-0001-8971-965X}{\usebox{\orcid}\hspace{1mm}Meiyazhagan Jaganathan}%
}
\author[1]{%
    \href{https://orcid.org/0000-0002-8318-3521}
    {\usebox{\orcid}\hspace{1mm}Vikram Pakrashi}%
}

\author[1]{%
	\href{https://orcid.org/0000-0003-1497-145X}{\usebox{\orcid}\hspace{1mm}Aasifa Rounak\thanks{corresponding author - \texttt{aasifa.rounak@ucd.ie}}}%
}

\affil[1]{UCD Centre for Mechanics, Dynamical Systems and Risk Laboratory, School of Mechanical and Materials
Engineering, University College Dublin, D04 V1W8, Dublin, Ireland}

\renewcommand{\headeright}{}
\renewcommand{\undertitle}{}
\renewcommand{\shorttitle}{Interplay of Nonsmoothness, Time Delay, and Stochasticity in Turning Dynamics}

\hypersetup{
pdftitle={Interplay of Non-Smoothness, Time Delay, and Stochasticity in Turning Dynamics},
pdfsubject={Meiyazhagan Jaganathan}
}

\begin{document}
\maketitle

\begin{abstract}
The stochastic dynamics of orthogonal metal cutting with both regenerative and nonsmooth frictional effects are investigated numerically in this paper. \textcolor{black}{The shortcomings of disregarding nonsmooth, here frictional, as well as stochastic effects in analyzing the stability of such a machining process are demonstrated.} Dynamics of the tool motion is observed to exhibit rich nonlinear phenomena such as stick-slip during chatter, with stochastic perturbations in cutting forces adding further complexity, leading to the occurrence of stochastic P and D bifurcations. Measures of entropy are found to be effective in quantifying the dynamical transitions occurring in the dynamics of the tool. Subsequently, basin stability analyses, modified to account for stochasticity and time-delays, are carried out to systematically investigate the dynamics of the cutting tool across multiple surface roughness profiles of the workpiece. Basin stability analyses indicate that chatter can be controlled by restricting initial tool displacement and controlling initial workpiece surface roughness, suggesting practical strategies to improve machining outcomes for precision manufacturing.
\end{abstract}

\vspace{2em}
\nomenclature[D]{$Y(t)$}{Tool displacement [m]}
\nomenclature[D]{$\dot{Y}(t)$}{Tool velocity [m\,s$^{-1}$]}
\nomenclature[D]{$\ddot{Y}(t)$}{Tool acceleration [m\,s$^{-2}$]}
\nomenclature[D]{$m$}{Tool mass [kg]}
\nomenclature[D]{$c$}{Damping coefficient [N\,s\,m$^{-1}$]}
\nomenclature[D]{$k$}{Stiffness [N\,m$^{-1}$]}
\nomenclature[D]{$D$}{Workpiece diameter [m]}
\nomenclature[D]{$R$}{Workpiece radius, $R = D/2$ [m]}
\nomenclature[D]{$N$}{Spindle speed [rev\,min$^{-1}$]}
\nomenclature[D]{$\mu$}{Friction coefficient [-]}
\nomenclature[D]{$\mu_s$}{Static friction coefficient [-]}
\nomenclature[D]{$\mu_d$}{Dynamic friction coefficient [-]}
\nomenclature[D]{$V_c$}{Cutting speed, $V_c=2\pi RN/60$ [m\,s$^{-1}$]}
\nomenclature[D]{$V_{ch}$}{Chip velocity [m\,s$^{-1}$]}
\nomenclature[D]{$V_\gamma$}{Relative chip velocity, $V_\gamma=V_{ch}-\dot{Y}\cos\gamma$ [m\,s$^{-1}$]}
\nomenclature[D]{$V_s$}{Stribeck velocity [m\,s$^{-1}$]}
\nomenclature[D]{$K$}{Cutting force coefficient [N\,m$^{-2}$]}
\nomenclature[D]{$a_p$}{Depth of cut [m]}
\nomenclature[D]{$C_y$}{Process damping coefficient [N\,s\,m$^{-2}$]}
\nomenclature[D]{$H_D$}{Nominal chip thickness (feed) [m]}
\nomenclature[D]{$H(t)$}{Instantaneous chip thickness [m]}
\nomenclature[D]{$F_\gamma$}{Tangential force on rake face [N]}
\nomenclature[D]{$F_{\gamma n}$}{Normal force on rake face [N]}
\nomenclature[D]{$F_r$}{Resultant force [N]}
\nomenclature[D]{$F_c$}{Cutting force ($x$-direction) [N]}
\nomenclature[D]{$F_f$}{Feed force ($y$-direction) [N]}
\nomenclature[D]{$F_d$}{Process damping force ($y$-direction) [N]}
\nomenclature[D]{$t_w$}{Rotational period of the workpiece $t_w=\frac{60}{N}$ [s] (time-delay)}


\nomenclature[N]{$\tau$}{Non-dimensional time, $\tau=t\sqrt{k/m}$ [-]}
\nomenclature[N]{$\tau_w$}{Non-dimensional delay, $\tau_w=60/n$ [-]}
\nomenclature[N]{$y(\tau)$}{Non-dimensional displacement, $y=Y/H_D$ [-]}
\nomenclature[N]{$\Delta \tau$}{Non-dimensional time-step [-]}
\nomenclature[N]{$y_1(\tau)$}{State 1, $y_1=Y/H_D$ [-]}
\nomenclature[N]{$y_2(\tau)$}{State 2, $y_2=\dot{y}_1$ [-]}
\nomenclature[N]{$\dot{y}(\tau)$}{Non-dimensional velocity [-]}
\nomenclature[N]{$n$}{Non-dimensional spindle speed, $n = N\sqrt{m/k}$ [-]}
\nomenclature[N]{$c_y$}{Non-dimensional process damping, $c_y=30C_y/(\pi RK)$ [-]}
\nomenclature[N]{$\mu_{vs}$}{Non-dimensional friction coefficient [-]}
\nomenclature[N]{$v_s$}{Non-dimensional Stribeck velocity [-]}
\nomenclature[N]{$\nu$}{Non-dimensional chip thickness parameter [-]}
\nomenclature[N]{$v_\gamma$}{Non-dimensional relative velocity [-]}
\nomenclature[N]{$\xi$}{Damping ratio, $\xi = c/\sqrt{mk}$ [-]}
\nomenclature[N]{$g$}{Friction law argument [-]}
\nomenclature[N]{$\Omega$}{Non-dimensional frequency, $\Omega = 2\pi/\tau_w$ [-]}
\nomenclature[N]{$\omega$}{Non-dimensional excitation frequency [-]}
\nomenclature[N]{$W$}{Non-dimensional depth parameter, $W = a_pK/k$ [-]}
\nomenclature[N]{$h(\tau)$}{Non-dimensional chip thickness, $h=1-y+y(\tau-\tau_w)$ [-]}
\nomenclature[N]{$f_f$}{Non-dimensional cutting force [-]}
\nomenclature[N]{$f_d$}{Non-dimensional damping force [-]}

\nomenclature[G]{$\gamma$}{Rake angle [degrees]}
\nomenclature[G]{$\phi$}{Shear angle [degrees]}

\nomenclature[S]{$\eta$}{Noise intensity [-]}
\nomenclature[S]{$\lambda_{OU}(\tau)$}{Ornstein-Uhlenbeck process [-]}
\nomenclature[S]{$\mu_{OU}$}{OU process mean [-]}
\nomenclature[S]{$\sigma$}{OU volatility [-]}
\nomenclature[S]{$\theta$}{OU mean reversion speed [s$^{-1}$]}
\nomenclature[S]{$dW(\tau)$}{Wiener process increments [-]}
\nomenclature[S]{$\mathbb{R}^n$}{Euclidean state space of dimension $n$ [-]}
\nomenclature[S]{$\mathbf{X}_t$}{State vector, $\mathbf{X}_t \in \mathbb{R}^n$ [-]}
\nomenclature[S]{$\mathbf{f}(t,\mathbf{X}_t)$}{Drift term (deterministic dynamics) [-]}
\nomenclature[S]{$\mathbf{g}(t,\mathbf{X}_t)$}{Diffusion term (stochastic magnitude) [-]}
\nomenclature[S]{$\sigma_g$}{Noise intensity (stochastic scaling) [-]}
\nomenclature[S]{$dt$}{Deterministic time increment [-]}
\nomenclature[S]{$\mathbf{X}_0$}{Initial state vector, $\mathbf{X}_0 \in \mathbb{R}^n$ [-]}
\nomenclature[S]{$A_S$}{Attractor set [-]}
\nomenclature[S]{$B(\mathbf{X}_0)$}{Basin of attraction of initial state [-]}
\nomenclature[S]{$P(\cdot)$}{Probability measure [-]}
\nomenclature[S]{$\Lambda$}{Lyapunov Exponent [-]}
\nomenclature[S]{$y_{pert}(\tau)$}{Perturbed trajectory [-]}
\nomenclature[S]{$\delta$}{Small perturbation}
\nomenclature[S]{$\mathbb{E}$}{Expectation operator}
\nomenclature[S]{$T_s$}{\textcolor{black}{Duration of the stationary window}}


\nomenclature[C]{$W^*$}{Fixed Non-dimensional depth of cut value [-]}
\nomenclature[C]{$\|F\|_{\text{tol}}$}{Residual tolerance [-]}
\nomenclature[C]{$\text{dup}_{\text{tol}}$}{Duplicate tolerance [-]}
\nomenclature[C]{$\text{inc}$}{Predictor step size [-]}
\nomenclature[C]{$\text{rat}$}{Continuation ratio [-]}
\nomenclature[C]{$\text{maxC}$}{Max continuation points [-]}
\nomenclature[C]{$\text{maxT}$}{Max second-point attempts [-]}
\nomenclature[C]{$\bar{n}$}{Spindle speed search grid [-]}
\nomenclature[C]{$\bar{\omega}$}{Frequency search grid [-]}

\nomenclature[E]{$H_{SHE}$}{Shannon entropy}
\nomenclature[E]{$H_{CE}$}{Conditional entropy}
\nomenclature[E]{$H_{PE}$}{Permutation entropy}
\nomenclature[E]{$H_{AE}$}{Approximate entropy}
\nomenclature[E]{$H_{SE}$}{Sample entropy}
\nomenclature[E]{$H_{FE}$}{Fuzzy entropy}


\nomenclature[A]{CNC}{Computer Numerical Control}
\nomenclature[A]{DOF}{Degree(s) of Freedom}
\nomenclature[A]{DDE}{Delay Differential Equation}
\nomenclature[A]{\textit{j-pdf}}{The joint probability density}
\nomenclature[A]{OU}{Ornstein-Uhlenbeck process}
\nomenclature[A]{P-bifurcation}{Phenomenological Bifurcation}
\nomenclature[A]{D-bifurcation}{Dynamical Bifurcation}
\nomenclature[A]{LE}{Lyapunov Exponents}
\nomenclature[A]{RMS}{\textcolor{black}{Root Mean Square}}
\nomenclature[A]{SDDE}{\textcolor{black}{Stochastic Delay Differential Equation}}
\nomenclature[A]{SLD}{\textcolor{black}{Stability Lobe Diagram}}
\nomenclature[A]{RPM}{\textcolor{black}{Revolutions Per Minute}}
\printnomenclature[5cm]


\nomenclature[B]{$s(\theta_{bs}(\tau))$}{Initial workpiece waviness profile [-]}
\nomenclature[B]{$p_u(\theta_{bs}(\tau))$}{Uncut chip thickness profile [-]}
\nomenclature[B]{$\theta_{bs}(\tau)$}{Tool angular position [rad]}
\nomenclature[B]{$N_{bs}$}{Number of Fourier harmonics [-]}
\nomenclature[B]{$a_i, b_i$}{Fourier coefficients of waviness [-]}
\nomenclature[B]{$\phi_{bs}$}{Initial engagement phase angle [rad]}
\nomenclature[B]{$h_w$}{Waviness height [mm]}
\nomenclature[B]{$l_w$}{Waviness length [mm]}
\nomenclature[B]{$\alpha$}{Waviness amplitude constraint [-]}
\nomenclature[B]{$\beta$}{Initial vibration amplitude constraint [-]}
\nomenclature[B]{$M_{bs}$}{Number of Monte Carlo samples [-]}
\nomenclature[B]{$M_{CH}$}{Number of samples leading to chatter [-]}
\nomenclature[B]{$R_a$}{\textcolor{black}{Average roughness [-]}}
\nomenclature[B]{$R_q$}{\textcolor{black}{Root mean square roughness [-]}}
\nomenclature[B]{$M_{sto}$}{\textcolor{black}{Number of stochastic Monte Carlo runs [-]}}

\nomenclature[L]{$y_{10}$}{Equilibrium tool displacement (steady-state cutting) [-]}
\nomenclature[L]{$\boldsymbol{A}$}{Instantaneous system matrix [-]}
\nomenclature[L]{$\boldsymbol{D}$}{Delayed system matrix [-]}
\nomenclature[L]{$a,\, b$}{Linearized cutting-force coefficients (see Eq.~\eqref{eq-coeffi}) [-]}
\nomenclature[L]{$\lambda$}{Eigenvalue of the linearized delay system [-]}
\nomenclature[L]{$\omega$}{Hopf frequency at the stability boundary [rad/s]}
\nomenclature[L]{$\boldsymbol{I}$}{Identity matrix [-]}
\nomenclature[L]{$\det(\bullet)$}{Determinant operator [-]}


\section{Introduction}
\noindent Nonsmooth dynamical systems are essential for modeling physical processes with discontinuities, impacts, or switching events \cite{bernardo2008piecewise}. In mechanical systems, these phenomena often result from contact \cite{popp2000non}, friction, and intermittent engagement \cite{chin1994grazing}, causing abrupt transitions in forces and system responses. Metal cutting is one such important area of application where nonsmooth effects become dominant due to frictional effects or when the tool periodically establishes contact with the workpiece and disengages contact after a certain amount of time, leading to self-excited vibrations, such as chatter \cite{wiercigroch2006applied}. Chatter, the back-and-forth vibration of a tool against the workpiece surface, is a major problem for stable machining. It causes material waste, reduced surface quality, reduces tool life through wear and damage, exacerbates safety issues, including the generation of high-decibel noise, lowers productivity, and increases the cost of cutting processes. Despite advancements in Computer Numerical Control (CNC) technology, machine operations still face challenges from chatter \cite{amrc_chatter2023}. 

Chatter is classified into four categories based on the source of the cutting instability: (i)~frictional \cite{turning-wiercigroch2001frictional, turning-rusinek2014modelling}, (ii)~regenerative \cite{turning-wahi2008self, turning-stepan2001modelling}, (iii)~mode-coupling \cite{review2024zhang}, and (iv)~thermo-mechanical \cite{wiercigroch2001sources}. Among these, frictional and regenerative chatter have attracted significant attention in both nonlinear dynamics and manufacturing communities due to their frequent occurrence and most substantial influence on machining stability. Frictional chatter, also known as primary chatter, is driven by the velocity-dependent frictional effects at the interface between the rake face of the tool and the chip. Regenerative chatter, the secondary type, arises from waviness in the workpiece geometry resulting from prior passes, which affects the dynamics of subsequent passes, leading to higher cutting forces and dynamical instabilities \cite{review2024zhang}.  Orthogonal metal cutting is susceptible to both frictional and regenerative effects. 

\par Orthogonal metal cutting is an extensively used machining technique for manufacturing, where a single-point cutting tool, oriented orthogonally with the workpiece, generates \textit{chips} by removing material through shear deformation. Understanding and controlling the dynamics involved in orthogonal cutting, such as \textit{turning} operations, is essential for improving machining efficiency and precision. Turning operations have been extensively studied in the past decades. Wiercigroch {\it et al.} \cite{wiercigroch2001sources} revisited classical and modern models of \textit{chip formation} and cutting mechanics, emphasizing the importance of nonlinearities in understanding and predicting machining behavior. A two-degree-of-freedom (2-DOF) frictional model for the machine-tool and cutting process system was introduced in 2001~\cite{turning-wiercigroch2001frictional}, where the authors explored the emergence of period-doubling bifurcations and chaotic dynamics in orthogonal metal cutting. In 2014, Rusinek \textit{et al.}\cite{turning-rusinek2014modelling} improved the 2-DOF frictional model by incorporating the flank forces, demonstrating how  \textit{flank forces} play a major role in stable dynamics. G\'abor St\'ep\'an's 2001 work \cite{turning-stepan2001modelling} introduced a time-delay nonlinear model for regenerative turning, addressing the limitations of linear models and proposing a three-quarter power law for cutting forces. Investigation of the global dynamics of regenerative chatter in turning in the deterministic sense was presented by Wahi \textit{et al.} ~\cite{turning-wahi2008self}, with particular focus on the phenomenon of self-interrupted cutting due to contact loss between tool and workpiece. A single-degree-of-freedom (SDOF) nonsmooth delay differential-algebraic equation was later introduced by Dombovari \textit{et al.} ~\cite{turning-dombovari2011global} to analyze the chatter in regenerative orthogonal cutting. This study also employed DDE-BIFTOOL \cite{engelborghs2001dde} to perform continuation analysis on the smooth version of the model. 

\par \textit{Chip formation} in orthogonal cutting has been studied for titanium alloys such as (Ti6Al4V) \cite{turning-bai2017improved}, and the influence of chip segmentation on the cutting stability has also been explored \cite{turning-gyebroszki2018stability}. Additionally, the role of \textit{process damping} in orthogonal cutting has received significant attention. Ahmadi \cite{turning-ahmadi2017analytical} analytically modeled the process damping arising from flank contact. Eynian and Altintas \cite{turning-Eynian_chatter_stability} developed two models, with their advanced formulation explicitly modeling the process damping from flank wear, showing its stabilising effect. Similarly, Yuvaraju \textit{et al.} \cite{turning-yuvaraju2023nonlinear} introduced a 2-DOF nonlinear model for internal turning that includes flank-induced damping and frictional effects, capturing complex dynamics like internal resonance. Investigations into the chatter phenomenon have extended beyond turning and orthogonal cutting to encompass a wide range of machining processes. In \textit{grinding}, studies have explored the onset and suppression of chatter under various grinding conditions \cite{grinding-altintas2004chatter, grinding-li2007studygrind, grinding-yan2014chattergrind, grinding-yan2018stabilitygrind, grinding-liu2025chatter}. In \textit{drilling}, generalized models have been developed to study chatter stability \cite{drilling-roukema2007generalized-pt1, drilling-roukema2007generalized-pt2, drilling-stone2004stability}. \textit{Boring} operations have also been analyzed for chatter stability, particularly in robotic boring systems \cite{boring-wang2017chatter}. \textit{Milling}, one of the most extensively studied cutting processes, has seen the development of time-domain and frequency-domain models \cite{milling-altintas2008chatter, milling-insperger2003multiple, milling-zhu2020recent}.  Additional complexity like chatter stability (\cite{altintas2020chatter}) is introduced for \textit{Parallel Machining}, where multiple tools cut simultaneously.  Further insights into these various investigations and a more comprehensive history of studies around the phenomenon of chatter can be found in several review articles \cite{review2024zhang, review2011quintanachatter, review2012siddhpura, reviewsujuan2021analytical}. 

\par Most of these studies, representing a wide range of models exhibiting chatter, rely on delay differential equations (DDEs) based deterministic models where the machining parameters, including cutting force coefficients, typically rely on averaged constant values. As the majority of machining models in use at present are deterministic, insights obtained from such models are approximate, conservative, and even correspond to rules of thumb where stochastic aspects are relevant. The dynamic behavior and stability of tools in real machining operations are influenced by unmodeled stochastic factors, such as inhomogeneities in material quality, chip discontinuity, the workpiece surface, randomness in the cutting resistance, and friction \cite{turning-wiercigroch2001frictional, turning-noise-fodor2020}, or due to fluctuations during machining. The combined stochastic effect of such unmodeled effects from disparate sources can be incorporated into the governing equations as \textit{noise}. In this context, noise captures random variations in cutting force parameters of the DDE. Noise can alter the dynamics and cause the system to intermittently hop between the underlying deterministic multiple stable states, leading to noise-induced intermittency \cite{kumar2017bifurcation}. This is distinct from other routes to intermittency known in the context of deterministic systems\cite{hilborn2000chaos}. As notions of stable states, bifurcations, and routes to intermittency in the stochastic context are different, and such systems require separate analyses \cite{arnold1996toward}. The influence of noise in amplification of machine tool vibrations due to coherence resonance has been studied, along with effects on machine tool dynamics ~\cite{turning-noise-buckwar2006noise}. Stochastic analyses of regenerative turning models ~\cite{turning-noise-fodor2020, turning-noise-gradivsek2002stochastic, turning-noise-sykora2017theoretical} highlight how noisy fluctuations alter the stability characteristics of turning processes, while a recent work \cite{turning-noise-fodor2024efficient} utilizes power spectral density to approximate the dynamical states in a stochastic turning process. While these works separately focus on the regenerative or frictional effects of the orthogonal cutting processes in varied detail, it is also possible to approximate the dynamics of the cutting tool to a fundamental SDOF model with both effects combined \cite{turning-yan2019modelling}. This is a more realistic scenario of what happens during machining, allowing for investigations into 
nonlinear stability and improving the accuracy of cutting stability prediction, particularly in low-velocity machining \cite{turning-yan2021safety}.  

\par It is observed that, irrespective of the variety and complexity of investigations present in this field, the combined impact of regenerative and frictional nonlinearities under stochastic fluctuations remains unexplored. This omission is critical, as real machining processes are inherently subject to uncertainties arising from material inhomogeneities, chip discontinuities, and tool-workpiece interactions, all of which can significantly alter system stability. To address this gap, the present study builds upon the model proposed by Yan {\it et al.} \cite{turning-yan2019modelling} and introduces stochasticity in the DDE model, including both regenerative and nonsmooth frictional effects. Incorporating noise into cutting force dynamics allows for the representation of realistic operating conditions and subsequent investigation of the influence of such random fluctuations within the multi-stable region, particularly on transitions between coexisting states for the first time.


\par The rest of this paper is organized as follows. In Section~\ref{sec2}, the mathematical formulations of both the deterministic and stochastic DDE models for turning are introduced. Section~\ref{sec3} presents a linear stability analysis and discusses the role of nonsmooth friction in predicting stability. It also provides a nonlinear bifurcation analysis of the deterministic model, identifying three distinct stability regions. In Section~\ref{sec4}, stochastic bifurcations, namely, P- and D-bifurcations, are discussed. In Section~\ref{sec5}, the stability of the stochastic attractors is studied by computing the basin stability and comparing the values of various entropy measures. Finally, the outcomes of the work are summarized in Section~\ref{conclusion}.


\section{Mathematical Model}\label{sec2}
\begin{figure}[!hbt]
    \centering
    \includegraphics[width=.7\linewidth]{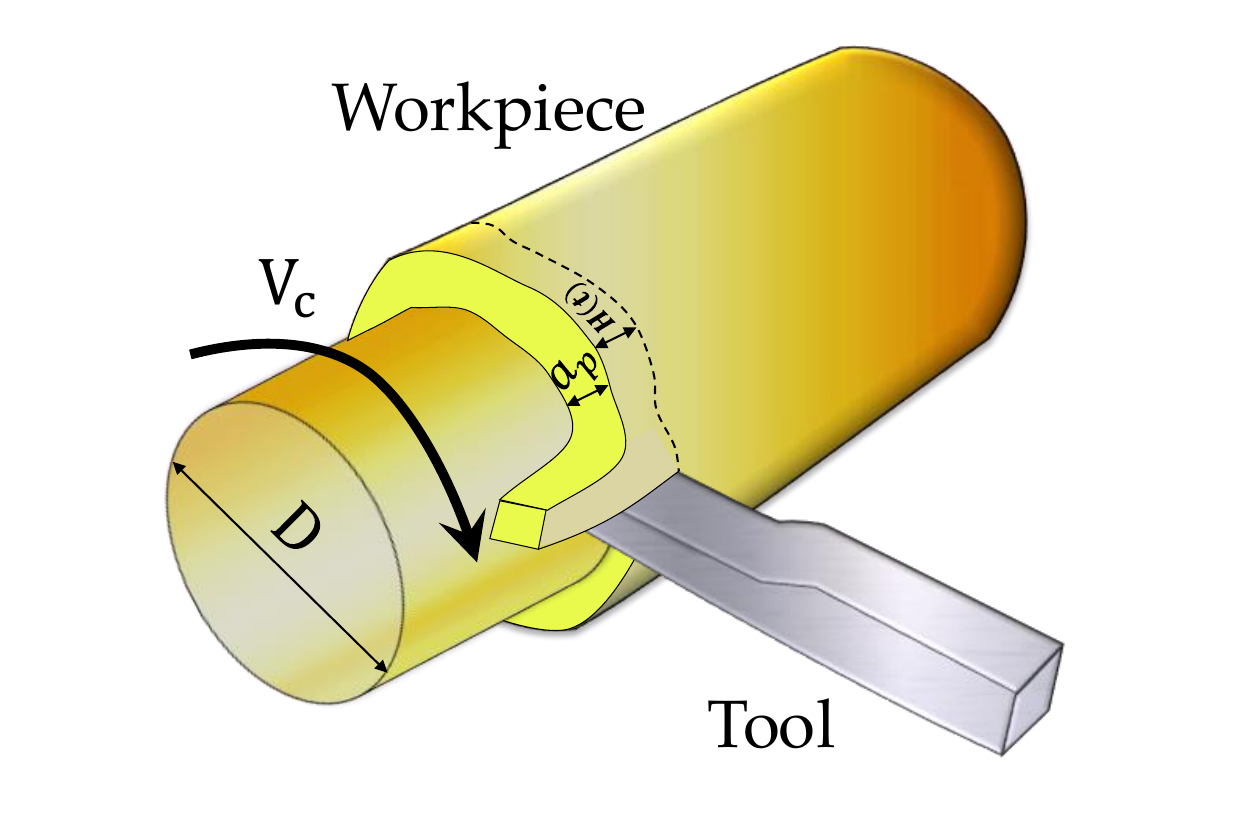}
    \caption{Schematic of the turning process representing the orientation of the tool and workpiece and their relative directions of motion.}
    \label{fig:Turning-1}
\end{figure}

\noindent The cutting process, illustrated in Fig.~\ref{fig:Turning-1}, refers to the complex dynamical interactions between a machine tool and a workpiece. To systematically analyze the dynamics of the tool, a deterministic model is first introduced, followed by the incorporation of multiplicative stochastic effects to account for unmodeled effects.

\subsection{The Deterministic Machining Model}
\begin{figure}[!hbt]
    \centering
    \includegraphics[width=.7\linewidth]{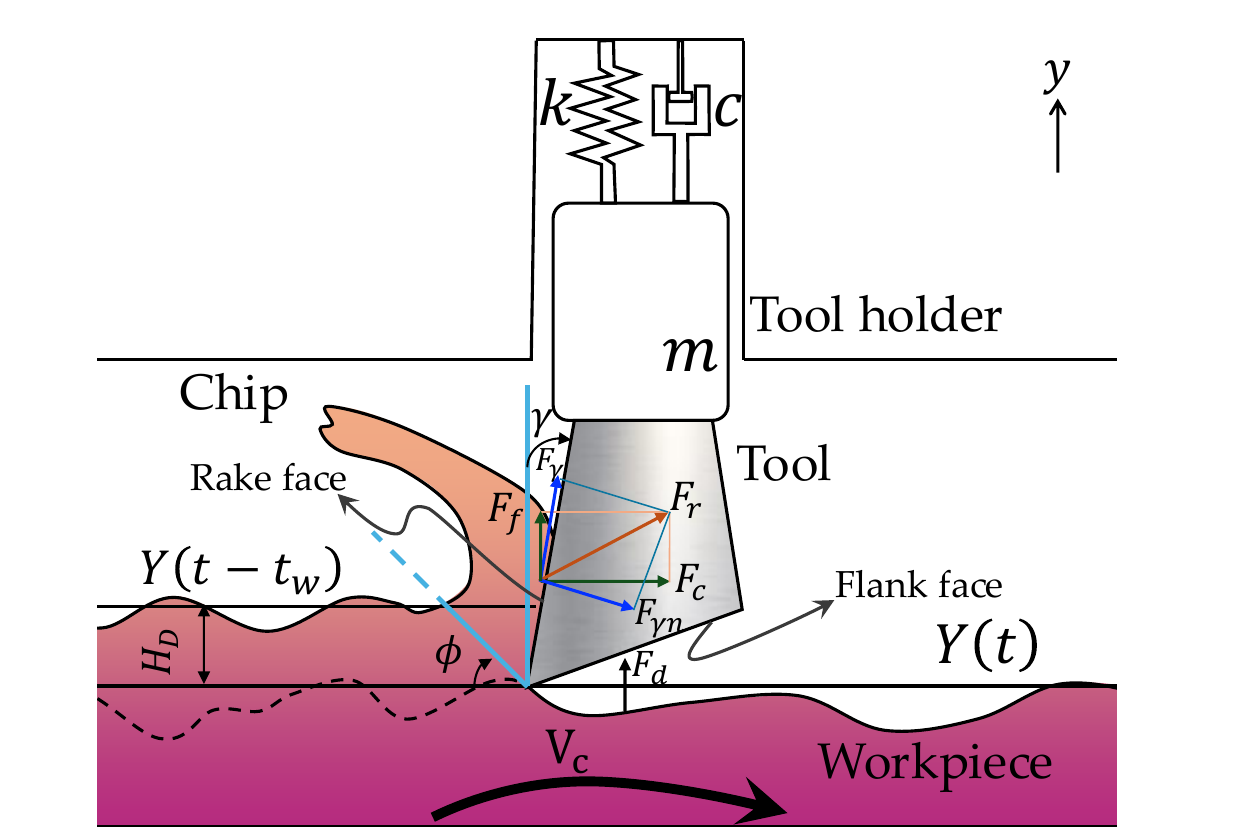}
    \caption{Schematic of the turning process modeled as a spring-mass-damper system representing various cutting forces.}
    \label{fig:Turning-2}
\end{figure}

\noindent Tool's dynamics in a turning process is modeled as a classical single degree of freedom (SDOF) model with damping $c\;[N\;s\;m^{-1}]$, stiffness $k\;[N\;m^{-1}]$, and mass $m\;[kg]$ corresponding to the tool \cite{turning-yan2019modelling} as shown in the schematic in Fig.~\ref{fig:Turning-2}. The radius of the workpiece is $R=D/2\;[m]$ and it has the workpiece speed of $V_c=\frac{2\pi R N}{60} [ms^{-1}]$ where $N\;[rev\;min^{-1}]$ here represents the spindle speed. During turning, the chip is formed through the process of yielding elasto-plastic deformation when the stationary tool and the rotating workpiece make contact \cite{wiercigroch2001sources}. 


During cutting, the material flows along the tool's rake face, which is inclined at an angle $\gamma$ and oriented perpendicular to the workpiece, with a velocity of $V_{ch}\;[ms^{-1}]$. In Fig.~\ref{fig:Turning-2}, the forces $F_\gamma\;[N]$ and $F_{\gamma n}\;[N]$ represent the tangential and normal components generated by the pressing and rubbing of the material against the rake face. The resultant force $F_r\;[N]$ is then resolved into the cutting force $F_c\;[N]$ in $x$-direction and feeding force $F_f\;[N]$ in $y$-direction. The interaction between the uncut material and flank face of the tool results in process damping force $F_d\;[N]$ in the $y$-direction. The dynamics of the tool shown in Fig.~\ref{fig:Turning-2} in the vertical direction is governed by the equation
\begin{equation}
    m\Ddot{Y}(t)+c\dot{Y}(t)+kY(t) = F_f + F_d. \label{eq-dim}
\end{equation}
The expressions for the forces are given by,
\begin{subequations}
    \begin{equation}
        F_d = - C_y a_p \dfrac{\dot{Y}(t)}{V_c},
    \end{equation}
    \begin{equation}
        F_f = F_\gamma \cos{(\gamma)} - F_{\gamma n} \sin{(\gamma)},
    \end{equation}
\end{subequations}
where $C_y$, $a_p$, and $\gamma$ denote the process damping coefficient, depth of cut, and rake angle of the tool, respectively.
The normal and tangential forces are of the form
\begin{subequations}\label{eqn-gamma-forces}
    \begin{equation}
        F_{\gamma n} = K a_p H(t),
    \end{equation}
    \begin{equation}
        F_{\gamma} = \mu F_{\gamma n},
    \end{equation}
\end{subequations}
where $H(t)$ denotes the instantaneous thickness of the chip and has the form
\begin{equation}
    H(t) = H_D - Y(t) + Y(t-t_w).
\end{equation}
Here $H_D\;[m]$ and $t_w=\frac{60}{N}\;[s]$ represent the nominal thickness of the chip (feed rate) and the rotational period of the workpiece, respectively. $K$ in Eq.~\eqref{eqn-gamma-forces} represents the cutting force coefficient. The term $\mu$ is the equivalent frictional coefficient corresponding to the Stribeck frictional model \cite{turning-bai2017improved}
\begin{equation}
    \mu = \mathrm{sign}(V_\gamma)\left(\mu_d + (\mu_d-\mu_s)\exp{\left(-\dfrac{|V_\gamma|}{V_s}\right)}\right),
\end{equation}
where $V_s\;[m\;s^{-1}]$ is the Stribeck velocity, $\mu_s$,  and  $\mu_d$ are the static and dynamic friction coefficients respectively. $V_\gamma\;[m\;s^{-1}]$ represents the frictional velocity of the chip relative to the tool in the form

\begin{figure}[htbp]
\centering
\includegraphics[width=0.65\textwidth]{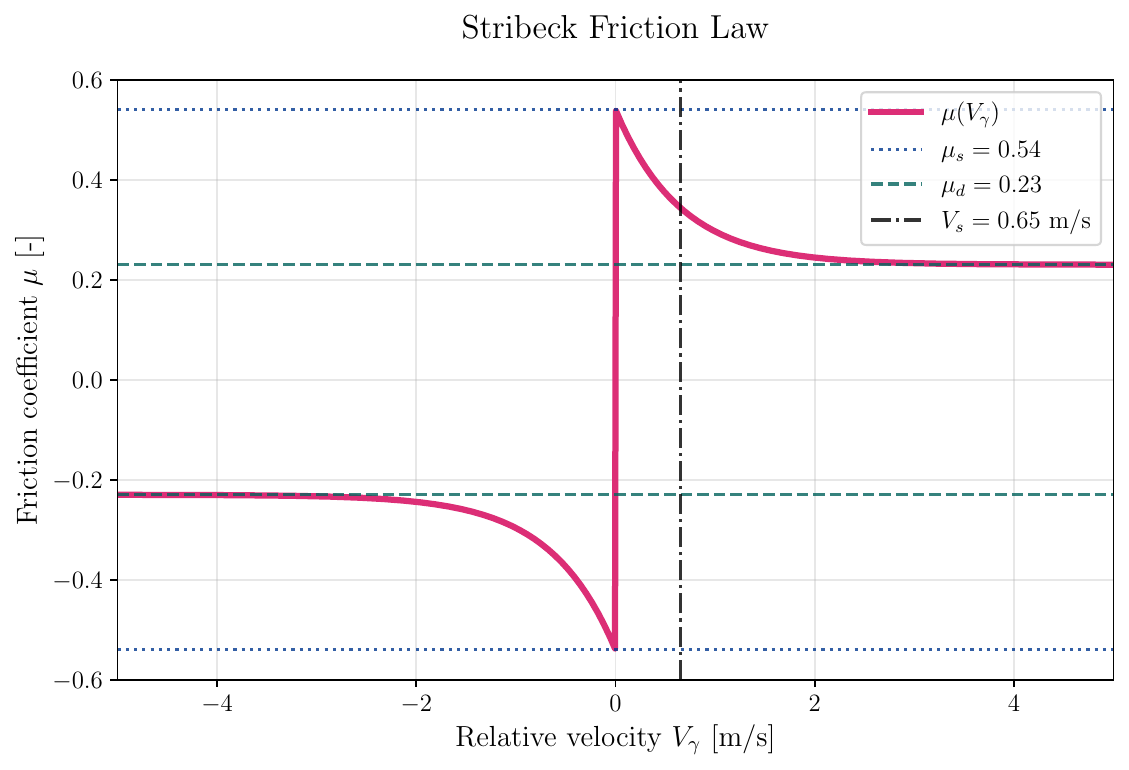}
\caption{Stribeck friction coefficient $\mu(V_\gamma)$ with parameters $\mu_d=0.23$, $\mu_s=0.54$, $V_s=0.65\;\mathrm{ms^{-1}}$. The antisymmetric curve shows an exponential transition from static to dynamic friction regimes around the Stribeck velocity.}
\label{fig:stribeck}
\end{figure}

\begin{equation}
    V_\gamma = V_{ch}-\dot{Y}(t)\cos(\gamma),
\end{equation}
where $V_{ch} = V_c\frac{\sin{(\phi)}}{\cos{(\phi-\gamma)}}\;[m\;s^{-1}]$ denotes the chip velocity and $\phi$ denotes the shear angle of the cutting tool. Fig.~\ref{fig:stribeck} illustrates the classical Stribeck friction law employed in the turning model. The friction coefficient exhibits characteristic stick-slip behavior, rapidly decaying from the static value $\mu_s$ at low velocities to the dynamic value $\mu_d$ for $|V_\gamma| \gg V_s$. The turning model described in Eq.~\ref{eq-dim} can be nondimensionalized as

\par $\xi=\dfrac{c}{\sqrt{mk}}$, $v_s=\dfrac{30V_s \cos{(\gamma-\phi)}}{\pi R \sin{(\phi)}}\sqrt{\dfrac{m}{k}}$, $c_y = \dfrac{30C_y}{\pi R K}$,
$\nu=\dfrac{H_D}{V_s}\sqrt{\dfrac{k}{m}}$, $n=N\sqrt{\dfrac{m}{k}}$, $\tau_w=\dfrac{60}{n}=t_w\sqrt{\dfrac{k}{m}}$, $\Omega=\dfrac{2\pi}{\tau_w}$, 

and
\begin{subequations}\label{non-dim-eqs}
    \begin{equation}
        W = a_p \dfrac{K}{k},  
    \end{equation}
    \begin{equation}
        \tau=t\sqrt{\dfrac{k}{m}},
    \end{equation}
    \begin{equation}\label{Eq-nondimY}
        y_1(\tau) = \dfrac{Y(t)}{H_D},
    \end{equation}
    \begin{equation}
        y_2(\tau)=\dot{y}_1(\tau)=\dfrac{\dot{Y}(t)}{H_D}\sqrt{\dfrac{m}{k}},
    \end{equation}
    \begin{equation}
        v_\gamma = \dfrac{V_\gamma}{H_D}\sqrt{\dfrac{m}{k}} = \dfrac{n}{v_s}-\nu \cos(\gamma) y_2(\tau).
    \end{equation}
\end{subequations}
The non-dimensional equation of the model is obtained as follows:

\begin{subequations}
    \begin{equation}
        \ddot{y}(\tau)+\xi \dot{y}(\tau) + y(\tau) = f_f +f_d, \label{eq-nondim}
    \end{equation}
with 
    \begin{equation}
        f_f = W (\mu_{vs}\cos{(\gamma)-\sin{(\gamma)}}) h(\tau), \label{eq:ff}
    \end{equation}
    \begin{equation}
        f_d = - W c_y \dfrac{\dot{y}}{n}, \label{eq:fd}
    \end{equation}
    \begin{equation}
        \mu_{vs} = \mathrm{sign}(g)\left(\mu_d+(\mu_s-\mu_d)e^{|-g|}\right),
    \end{equation}
    \begin{equation}
  \text{where} \,\,\,  g = \frac{n}{v_s(\gamma,\phi)}-\nu \cos (\gamma) \dot{y}(\tau),
    \end{equation}
    \begin{equation}
     \text{and} \, \,\,\,  h(\tau)=1-y(\tau)+y(\tau-\tau_w).
    \end{equation}
\end{subequations}


\subsection{The Stochastic Dimensionless Equation for Turning}
\noindent Turning involves complex interactions and unpredictable factors, making it a highly nonlinear and stochastic process. This is reflected in both qualitative and quantitative behaviors observed in deterministic numerical models that simulate such processes \cite{altintas2020chatter,turning-yan2019modelling,turning-dombovari2011global,turning-Eynian_chatter_stability}. \textcolor{black}{In practical turning operations, cutting force demonstrates limited repeatability, even when cutting conditions are nominally constant. Microscopic variations in material hardness, intermittent chip segmentation, local surface defects, and progressive tool wear generate random fluctuations in chip load and tool-chip friction. These factors manifest as the effective perturbations of the cutting forces. To capture the combined influence of these factors, stochasticity is incorporated multiplicatively in the depth-of-cut term of the cutting force. This modeling approach acknowledges that identical nominal cutting parameters can yield different instantaneous force levels from one revolution to the next, and even within a single revolution.} 
Here, a noise term is introduced to the expressions of the \textcolor{black}{non-dimensional} depth of cut in the cutting forces \eqref{eq:ff} and \eqref{eq:fd} to study the combined effects of these complex processes. The resulting equation now accommodates unmodeled effects such as workpiece material inhomogeneities, surface roughness on the workpiece or tool wear, built-up edge over the rake and flank faces, or environmental disturbances. The resulting non-dimensional Stochastic Delay Differential Equations (SDDE) is 
 \begin{equation}
        \ddot{y}(\tau)+\xi \dot{y}(\tau) + y(\tau) = \bar{f}_f +\bar{f}_d, \label{eq-nondim-stoch}
\end{equation}
with

\begin{subequations}\label{eq:noise-force}
    \begin{equation}
	\bar{f}_f = W (1+\eta \lambda_{OU}(\tau))( \mu_{vs} \cos \gamma - \sin \gamma) h(\tau),
    \end{equation}
    \begin{equation}
	 \bar{f}_d = -W(1+\eta \lambda_{OU}(\tau)) c_y \dfrac{\dot{y}}{n}, 
    \end{equation}
\end{subequations}
where $\eta$ is the intensity of the noise and $\lambda_{OU} (\tau)$ is modeled as an Ornstein-Uhlenbeck (OU) process, governed by
\begin{equation}
	d\lambda_{OU}(\tau) = \theta \{\mu_{OU} - \lambda_{OU}(\tau)\} d\tau + \sigma dW(\tau),
\end{equation}
where $\mu_{OU}$ is the process mean, $\sigma$ is the volatility, $\theta$ denotes the mean reversion speed and $dW(\tau)$ are the increments of a Wiener process. An Ornstein-Uhlenbeck process is chosen to model the fluctuating term in the system since it is a stationary Gaussian Markov process with zero mean. \textcolor{black}{In this formulation, the noise directly perturbs the effective chip load and process damping forces, consistent with the physical interpretation of fluctuating chip thickness and frictional contact. Alternative stochastic models are theoretically possible. A Wiener process (Brownian motion) would result in unbounded, non-stationary random walks in the cutting force, which does not align with the bounded force fluctuations observed around a nominal mean. Poisson or jump processes are appropriate when rare, discrete events such as tool breakage or large chip spalling are dominant, but they do not capture the persistent small-amplitude fluctuations typical of stable cutting, and are also not suitable for the continuous frictional case. In contrast, the OU process offers a minimal continuous-time model with bounded variance and adjustable correlation time, making it well-suited to represent aggregated cutting force uncertainties in turning.} OU process has been used in several studies to model regenerative chatter in metal cutting where delay and stochastic effects interact across multiple time scales \cite{kuske2010competition}, investigate stochastic P-bifurcations in nonlinear impact oscillators with soft barriers \cite{rounak2020stochastic}, and analyze vibro-impact systems driven by noise \cite{liu2018probabilistic}.


\section{Deterministic Analysis}\label{sec3}
\subsection{Linear Stability Analysis}\label{sld-deter}
In steady state cutting, the tool's dissipative and elastic responses equilibrate the cutting forces. Under these steady-state conditions, the tool's static deformation is determined by solving the equilibrium form of Eq.~\eqref{eq-nondim}. This equation can be rewritten as
\begin{subequations}\label{eq-nondim2}
    \begin{equation}
        \dot{y}_1(\tau) = y_2(\tau),
    \end{equation}
    \begin{equation}
        \dot{y}_2(\tau) = -y_1(\tau) - \xi y_2(\tau) + W [\mu_{vs} \cos(\gamma) - \sin(\gamma)]h(\tau) - Wc_y \dfrac{y_2(\tau)}{n}.
    \end{equation}
\end{subequations}
For steady state, with no vibration, the tool displacement is constant, velocity is zero and there will be no regeneration effect ($y_1(\tau) = y_1(\tau - \tau_w) = y_{10}$) 
\begin{equation}
    \therefore \left(y_1(\tau), y_2(\tau)\right) \equiv \left(y_{10}, 0\right),
\end{equation}
where 
\begin{equation}
    y_{10} = W\left[\mu_d + (\mu_s-\mu_d) \exp\left(-\dfrac{n}{v_s}\right)\right]\cos(\gamma) - W \sin(\gamma). \label{eq-equ_point}
\end{equation}
A perturbation of the stationary cutting process causes the tool to oscillate around the equilibrium point given in Eq.~\eqref{eq-equ_point}. The local stability of $\left(y_{10}, 0\right)$ is then evaluated by linearizing Eq.~\eqref{eq-nondim2}, which yields
\begin{equation}
    \boldsymbol{\dot{y}}(\tau) = \boldsymbol{A}\boldsymbol{y}(\tau) + \boldsymbol{D}\boldsymbol{y}(\tau-\tau_w), 
\end{equation}
where 
\begin{equation}
    \boldsymbol{y} = \begin{pmatrix}
y_1 \\
y_2 
\end{pmatrix}, 
\boldsymbol{A} = \begin{pmatrix}
0 & 1 \\
-(1+Wa) & -(\xi+Wb)
\end{pmatrix}, 
\boldsymbol{D} = \begin{pmatrix}
0 & 0 \\
Wa & 0
\end{pmatrix}\label{eq-matrices}
\end{equation}
The coefficients $a$ and $b$ in Eq.~\eqref{eq-matrices} as follows
\begin{subequations}\label{eq-coeffi}
    \begin{equation}
        a = \left(\mu_d+(\mu_s-\mu_d)\exp\left(-\dfrac{n}{v_s}\right)\right)\cos(\gamma) - \sin(\gamma),
    \end{equation}
    \begin{equation}
        b = \dfrac{c_y}{n} + (\mu_d-\mu_s) \gamma \cos^2(\gamma)\exp\left(-\dfrac{n}{v_s}\right).
    \end{equation}
\end{subequations}
\par The stability is then determined by the eigenvalues of the following characteristic equation.
\begin{equation}
    \det \left( \lambda\boldsymbol{I}- \boldsymbol{A}-\boldsymbol{D}e^{-\lambda \tau_w}\right) = 0, \label{eq-char}
\end{equation}
where $\det(\bullet)$ represents a determinant and $\lambda$ denotes the eigenvalues. For the equilibrium to remain stable, all eigenvalues of the linearized system must lie strictly in the left half of the complex plane, meaning each has a negative real part. If, however, even a single eigenvalue acquires a positive real part, the perturbation grows instead of diminishing, and the stationary cutting becomes unstable. At the boundary between these two regimes, the point of critical stability is where the system exhibits a pair of purely imaginary eigenvalues, $\lambda = \pm i\omega$. This pair signals the onset of self-excited oscillations, as the system transitions from decaying motion to sustained vibration \cite{yan2017basins}. Substituting this critical condition into Eq.~\eqref{eq-char} transforms the characteristic equation 
\begin{equation}
    \det \left( i\omega\boldsymbol{I}- \boldsymbol{A}-\boldsymbol{D}e^{-i\omega \tau_w}\right) = 0, 
\end{equation}
or
\begin{subequations}\label{eq-sldeqn}
    \begin{equation}
        \omega^2 - 1 - Wa[1-\cos(\omega\tau_w)] = 0,
    \end{equation}
    \begin{equation}
        \omega(\xi+W b) + Wa\sin(\omega \tau_w) = 0.
    \end{equation}
\end{subequations}
The detailed derivation of Eq.~\eqref{eq-sldeqn} is provided in \ref{app1}. As indicated in Eq.~\eqref{eq-coeffi}, both $a$ and $b$ depend on non-dimensionalised spindle speed $n=\dfrac{\tau_w}{60}$, which renders Eq.~\eqref{eq-sldeqn} transcendental and prevents a closed-form analytical solution. Therefore, the stability lobes are computed via a numerical eigenvalue analysis using the Newton-Raphson continuation scheme. The complete continuation algorithm is provided in \ref{app2}. All numerical hyperparameters used in the predictor-corrector scheme are summarized here for clarity. The initial predictor step ($\text{inc}$) used to generate the second point is set to $2\times10^{-5}$. The continuation ratio ($\text{rat}$), which controls the predictor step scaling, is set to $1.005$. Each curve is allowed to grow to a maximum of $\text{maxC}=1000$ points, while the number of attempts to locate a valid second point is limited to $\text{maxT}=30$. The nonlinear solver accepts a solution only when the residual norm satisfies $\|F\|_{\text{tol}}=10^{-10}$, and duplicate seed points are removed using a strict tolerance of $\text{dup}_{\text{tol}}=10^{-12}$. Finally, the fixed parameter $W^*$ is selected such that the resulting $a_p$ slightly exceeds the region of interest (in this study, 2 mm).
 
\begin{figure}[!hbt]
    \centering
    \includegraphics[width=1\linewidth]{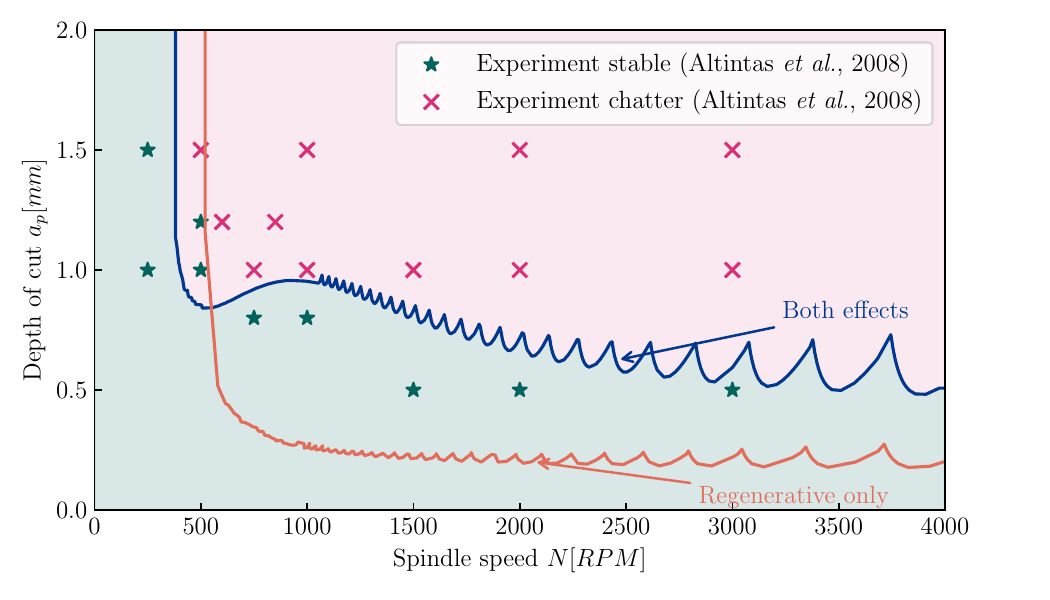}
    \caption{The stability lobe diagram shows the numerically computed stable region in green and the chatter region in magenta. The blue curve depicts the stability lobes for the model with both frictional and regenerative effects, whereas the orange curve shows the lobes for the regenerative and static-friction model. The red crosses denote chatter and the green stars indicate stable cutting as observed experimentally by Altintas \textit{et al.} \cite{altintas2008identification}.}
    \label{fig:sld2}
\end{figure}

For this entire paper, the parameter values are adopted from \cite{turning-yan2021safety}, with $m=0.561\;[kg]$, $c = 145\;[Nsm^{-1}]$, $k = 6.48\times 10^6\;[Nm^{-1}]$, $K = 6.02\times 10^9\;[Nm^{-2}]$, $C_y = 6.11\times 10^5\;[Nm^{-1}]$, $R = 0.0175 [m]$, $H_D = 0.0005\;[m]$, $\gamma = 0^o$, $\phi = 45^o$, $V_s = 0.65\;[ms^{-1}]$, $\mu_d = 0.23$, $\mu_s = 0.54$. These values yield the following dimensionless parameters: $\xi=0.07605$, $v_s=0.10436$, $\nu=2.61434$, and $c_y=0.05541$. 
By substituting these parameter values into Eq.~\eqref{eq-sldeqn}, the corresponding stability boundaries are computed. Fig.~\ref{fig:sld2} illustrates the stability lobe diagram (SLD) obtained from the numerical eigenvalue analysis. In Fig.~\ref{fig:sld2}, the blue line indicates the stability boundary computed from solving the critical equations \eqref{eq-sldeqn} of the model, which has both the regenerative and frictional effects. This line separates the space into two regions: the region below the curve is stable (shaded green), and the region above the curve is unstable (shaded magenta). The orange line denotes the stability lobes of the model in the absence of nonsmooth friction. In other words, the friction coefficient is taken as the static friction $\mu_{vs} = \mu_s$. This assumption shifts the stability boundary and the stable cutting data from experiments lie under the unstable region. Therefore, it is important to account for nonsmooth friction in the model to more accurately capture the system dynamics.

\subsection{Nonlinear Bifurcation Analysis}
\noindent Bifurcation analysis of the DDE model \ref{eq-nondim} is carried out by considering depth of cut $a_p\;[mm]$ as the bifurcation parameter for a fixed spindle speed of $N=3600~[rev\;min^{-1}]$. The DDE \eqref{eq-nondim} is numerically solved using MATLAB's  \textit{dde23} \cite{shampine2001solving} solver with the step size of $\Delta \tau =0.01$ for the initial deterministic study. The solver's tolerance settings are set to their default configuration: a relative tolerance (\textit{RelTol}) of \textit{$10^{-3}$} and an absolute tolerance (\textit{AbsTol}) of $10^{-6}$. For each value of the bifurcation parameter, the history function is initialized with the solution from the previous parameter value in both forward and backward simulations.  All simulations were performed using the non-dimensionalized form of the governing equations. However, for ease of physical interpretation, the results are reconstructed and presented in terms of dimensional quantities, specifically time, instantaneous chip thickness, and frictional chip velocity relative to the tool by applying Eq.~\eqref{non-dim-eqs}.

\par
\begin{figure*}[!hbt]
     \centering
          \includegraphics[width=\textwidth]{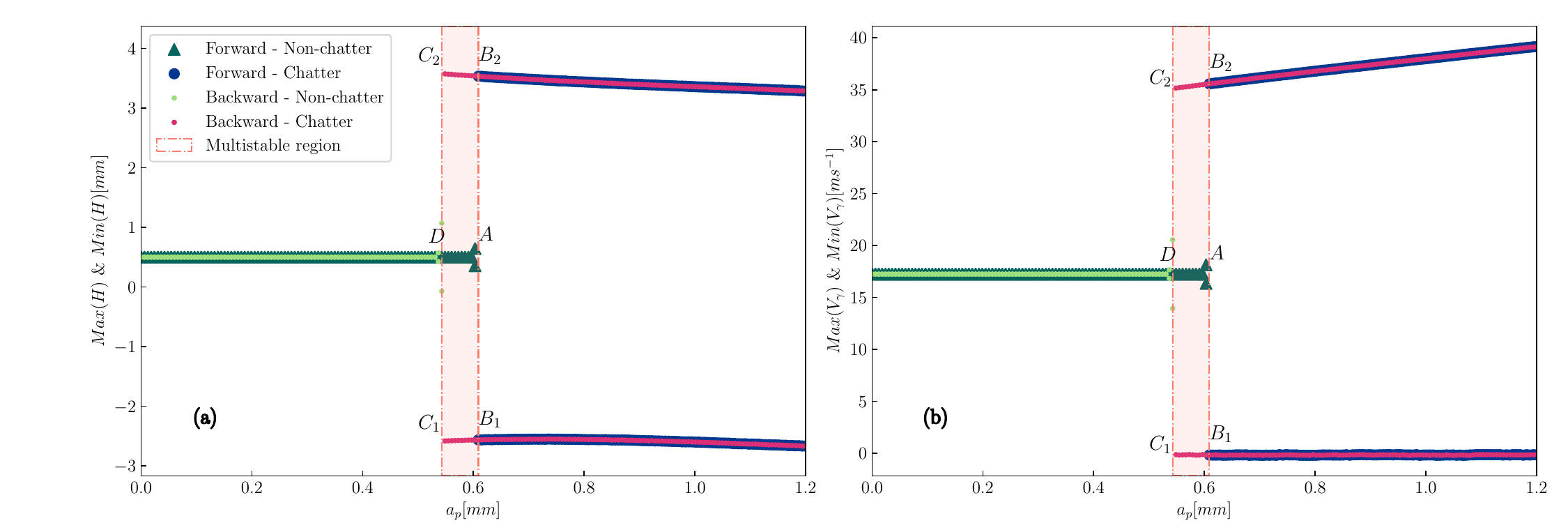}
         \caption{Bifurcation diagram using depth of cut $a_p$ as the bifurcation parameter. Figure (a) shows the maximum and minimum values of chip thickness $H$, and (b) shows the corresponding maximum and minimum values of frictional velocity $V_\gamma$.}
         \label{fig:bifur-deter}
\end{figure*}

 Fig.~\ref{fig:bifur-deter} shows the result of the bifurcation analysis where the maximum and minimum values of the instantaneous chip thickness $H(t)$ and the frictional chip velocity relative to the tool $V_\gamma$ in dimensional form are plotted along the $y$-axis while the depth of cut $a_p$ is in the $x$-direction. The bifurcation analysis reveals that the system's stability is divided into three distinct regions, as listed in Table~\ref{tab:attrac}.

\begin{table}[!hbt]
    \centering
    \begin{tabular}{|c|c|}
        \hline
         Region & Attractor\\ \hline
         Non-chatter & Fixed Point\\
         Chatter & Sticking limit cycle\\
         Multi-stable & Fixed Point and sticking limit cycle\\ \hline
    \end{tabular}
    \caption{List of regions with distinct phenomena and corresponding attractors in the dynamical systems sense.}
    \label{tab:attrac}
\end{table}

\begin{figure*}[!hbt]
     \centering
          \includegraphics[width=\textwidth]{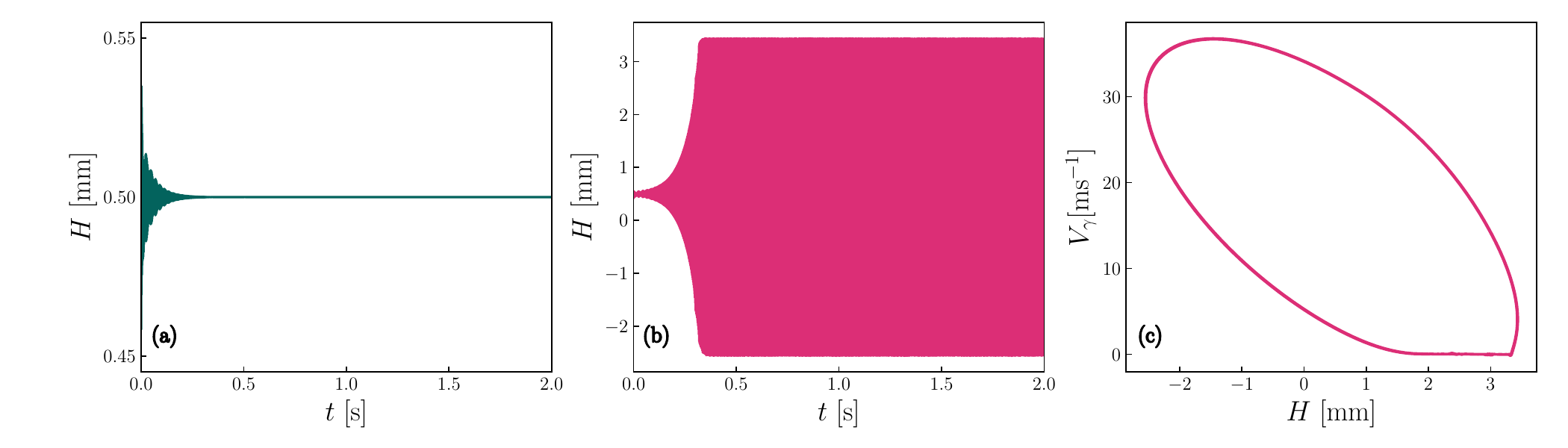}
         \caption{(a)~Variation of chip thickness over time for the stable case $a_p=0.4\;mm$, (b)~Variation of chip thickness over time for the unstable case $a_p=0.8\;mm$, and (c)~Phase portrait of an unstable cutting showing the sticking dynamics of the tool.}
         \label{fig:bifur-deter2}
\end{figure*}
The shaded region in Fig.~\ref{fig:bifur-deter} corresponds to the multi-stable region. The dynamical states to the left and right of the multi-stable region represent non-chatter and chatter regions, respectively. It can be observed that in the non-chatter region, the maximum and minimum values of $H(t)$ are the same, whereas in the chatter region, their magnitudes are different, indicating high-amplitude vibrations. Also, in the chatter region, the minimum values of $H(t)$ become negative, indicating loss of contact between the tool and the workpiece. This behavior typically arises from regenerative chatter, in which low spots created during the previous cutting pass influence the tool's engagement in the subsequent pass. The model captures this by dynamically updating the chip thickness $H(t)$ via a delay term that reflects the influence of past position on the tool's current position. This can also be observed from the Figs.~\ref{fig:bifur-deter2}~(a-b). The chip thickness variation over time is given in Fig.~\ref{fig:bifur-deter2}~(a) for $a_p=0.4\;mm$ and in Fig.~\ref{fig:bifur-deter2}~(b) for $a_p=0.8\;mm$. From Fig.~\ref{fig:bifur-deter2}~(a), it is observed that the chip thickness is constant after the transient, which makes the maximum and minimum values have the same value for $a_p=0.4\;mm$ in Fig.~\ref{fig:bifur-deter}. Similarly, Fig.~\ref{fig:bifur-deter2}~(b) shows the large amplitude vibrations which reflect two different magnitudes for maximum and minimum values for $a_p=0.8\;mm$ in Fig.~\ref{fig:bifur-deter}.  The phase portrait plotted between dynamic chip thickness and chip velocity for the chatter case is shown in Fig.~\ref{fig:bifur-deter2}~(c). From Figs.~\ref{fig:bifur-deter2}~(c) and \ref{fig:bifur-deter}~(b), one can also observe that the minimum value of frictional chip velocity relative to the tool $V_\gamma$ is zero during chatter cutting dynamics. This happens in an orthogonal cutting with a stationary cutting tool, when the chip velocity $V_{ch}$ matches the velocity of the tool surface at the rake face, where the rake face doesn't move. This alters the normal cutting mechanics, leading to chatter. The DDE \ref{eq-nondim} is integrated up to a dimensionless time of $\tau = 4.5 \times 10^{3} $, with the resulting dynamics subsequently represented in terms of the physical time variable $t$. Also, from Fig.~\ref{fig:bifur-deter} it is noticed that the cutting stability switches from point $A$ to points $B_1$ and $B_2$ during the forward simulation, corresponding to an increase in the depth of cut $a_p$ from $0.2\; mm$ to $1.2\;mm$. However, the results of backward simulation show that the stability jumps from points $C_1$ and $C_2$ to $D$, resulting in a multi-stable region at certain values of the bifurcation parameter. In this region, the system's stability depends significantly on the tool condition, the initial profile of the workpiece surface, and stochastic fluctuations, which account for various unmodeled effects. While the first two factors can typically be estimated and controlled before the cutting process begins, realistic scenarios across diverse application sectors underscore the critical need to understand system dynamics under noisy conditions. Incorporating noise into the numerical analysis is thus essential for clearly demarcating chatter occurrence, controlling chatter, and ensuring stable cutting throughout the cutting process.

\section{\textrm{Stochastic Analysis}}\label{sec4}
\noindent In deterministic systems, bifurcation analysis involves examining the system’s long-term behavior as parameters vary. However, stochastic systems are continuously influenced by external fluctuations, which prevent the dynamics from settling on the underlying deterministic attractors. In addition, there is always a non-zero probability of noise dislodging the dynamics from the vicinity of an attractor. As a result, the conventional definition of bifurcation used in deterministic settings must be adapted to account for these fluctuations when conducting stability analysis in stochastic systems. For a stochastic process given in Equation \ref{sp}
\begin{equation}\label{sp}
     \mathrm{d} \mathbf{X}_t=\mathbf{f}(t, \mathbf{X}_t) dt+\sigma_g \, \mathbf{g}(t, \mathbf{X}_t) d \mathbf{W}_t,\;\;\;\;\;\;\; \mathbf{X}_t \in \mathbb{R}^n,
\end{equation}
where $\mathbf{f}(t, \mathbf{X}_t)$ represents the deterministic component governing the system's dynamics, also known as the drift term, $\mathbf{g}(t, \mathbf{X}_t)$ represents the diffusion terms, governing the magnitude of stochastic fluctuations in the dynamical system, $\sigma_g$ is the noise intensity and $d\mathbf{W}_t$ is the Wiener process increment. For stochastic analysis, the stochastic DDE \eqref{eq-nondim-stoch} was solved using the Euler-Maruyama \cite{higham2021introduction} scheme in MATLAB. 

\begin{figure}[!hbt]
    \centering
    \includegraphics[width=0.8\linewidth]{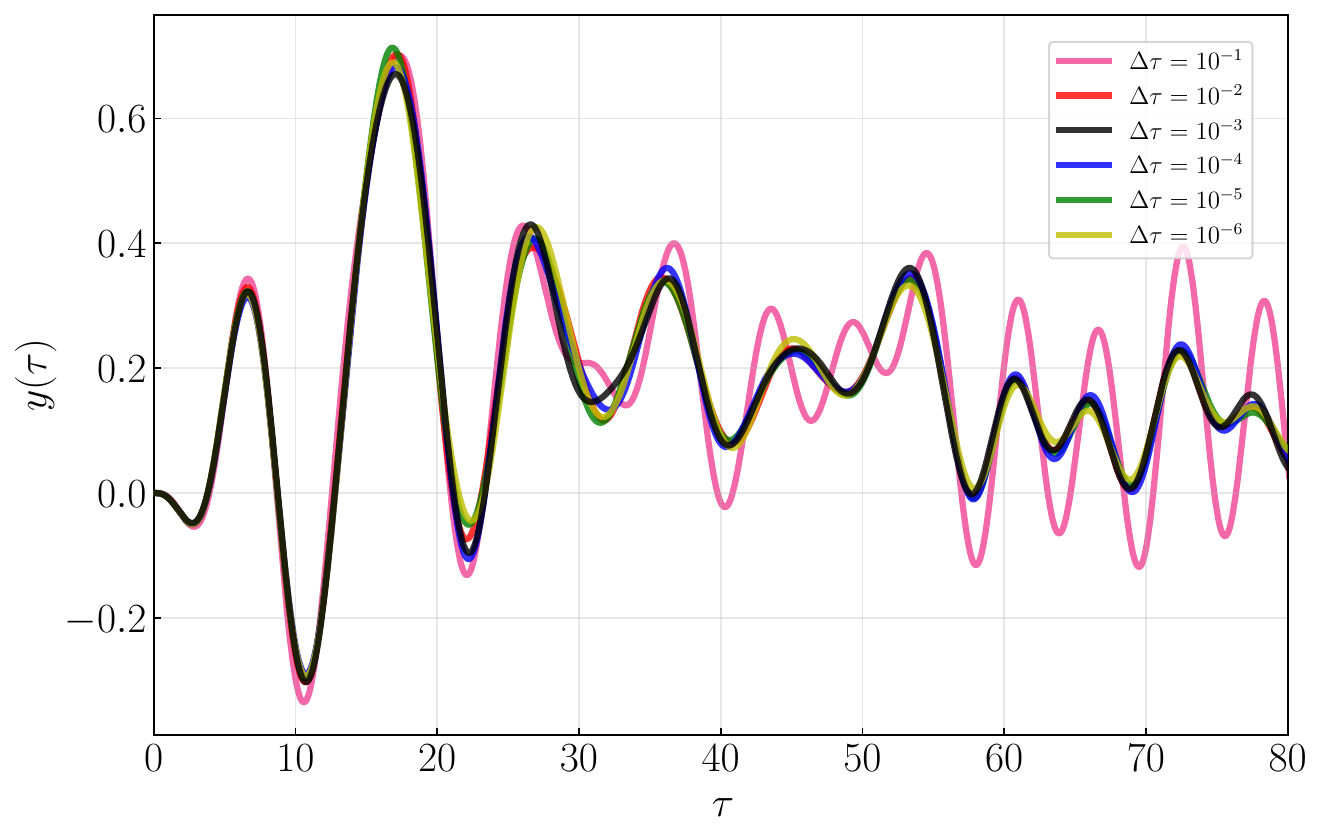}
    \caption{Comparison of numerical solution obtained with different time steps using the Euler-Maruyama scheme for the stochastic case. Pink, red, black, blue, green, and yellow color lines represent the numerical solution with time steps $10^{-1}$, $10^{-2}$, $10^{-3}$, $10^{-4}$, $10^{-5}$, and $10^{-6}$, respectively.}
    \label{fig:em_compar}
\end{figure}

The numerical solutions obtained with different step sizes $\Delta \tau$ from $10^{-1}$ to $10^{-5}$ validated against the solution at a very fine step size of ($\Delta \tau = 10^{-6}$), as illustrated in Fig.~\ref{fig:em_compar}. Analyzing peak-to-peak values of $y(\tau)$ reveals the magnitude of the optimal step size to be at $\Delta \tau = 10^{-3}$ with only 1.0\% relative amplitude error. This represents a substantial improvement over the coarser steps (2.3\% at $\Delta \tau = 10^{-2}$, 5.3\% at $\Delta \tau = 10^{-1}$) and thus was selected for further stochastic analysis simulations to ensure computational efficiency. The history function for this case is considered in the form of a Fourier expansion of multiple harmonics. More details regarding the history function and its relation with the initial configuration are given in section \ref{sec-basin}.

\begin{figure}[!hbt]
    \centering
    \includegraphics[width=1\linewidth]{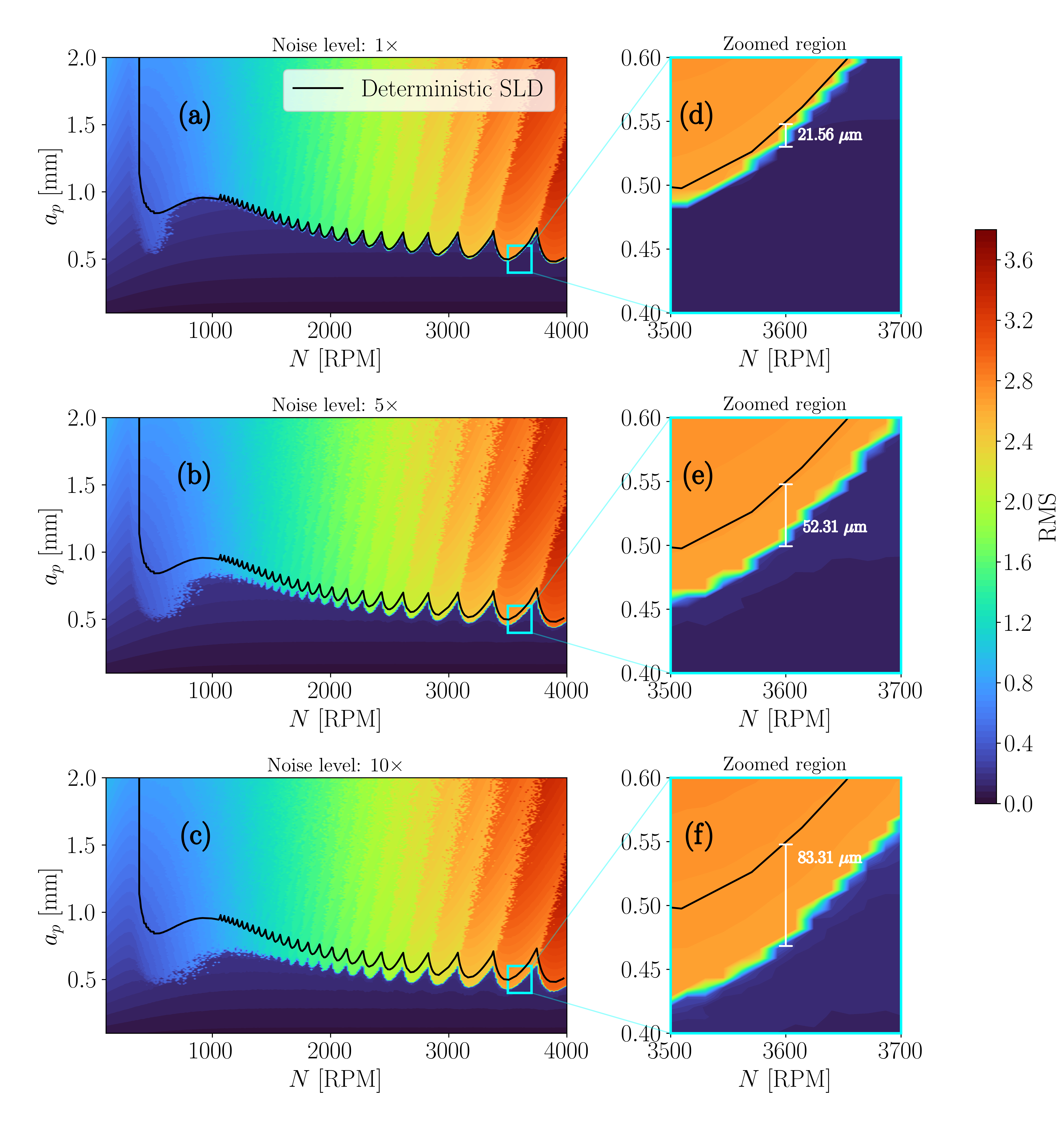}
    \caption{\textcolor{black}{Stationary RMS value of the nondimensional tool displacement of the model \eqref{eq-nondim-stoch} with (a) $1\times$  (b) $5\times$ and (c) $10\times$ noise amplifications respectively. Sub-figures (d)-(f) are the zoomed region of (a)-(c), respectively. The solid black represents the deterministic SLD, and the vertical white lines in panels (d)-(f) indicate the offset between the deterministic stability boundary and the RMS value corresponding to a $40\%$ nondimensional feed rate of $1$ at $N = 3600~RPM$.}}
    \label{fig:SLD noise}
\end{figure}

\textcolor{black}{Fig.~\ref{fig:SLD noise} shows the stationary Root Mean Square (RMS) measure of the tool displacement on $(N, a_p)$ parameter space for various amplified noise strength values. The value of the second moment is analyzed to assess the stability of the cutting process in Ref.~\cite{turning-noise-fodor2020}, where the authors use linear stochastic semi-discretization techniques to provide the second-moment stability analysis. The present model combines regenerative delay, velocity-dependent nonlinear friction, and OU-driven multiplicative noise, yielding a nonsmooth nonlinear SDDE. Semi-discretization is excellent for linear SDDE models, but the main phenomena of interest (nonsmooth friction, multi-stability, stochastic bifurcations) are fundamentally strongly nonlinear. Thus, a linearized SDDE, together with semi-discretization, would not faithfully capture these effects. Instead of linearizing around the steady cutting state, the full nonlinear SDDE is integrated numerically using the Euler-Maruyama algorithm. For each parameter pair $(N, a_p)$, long-time simulations are performed, and stationary vibration levels are estimated by time-averaged RMS measures of the tool displacement. Under the assumption of ergodicity in the stable regimes, these time averages approximate the stationary second moment, which is sufficient for assessing practical machining performance. Under the assumption of mean-square stability and ergodicity, the squared RMS approximates the stationary second-moment that would be obtained from stochastic semi-discretization for a linearized SDDE. The deterministic stability lobes obtained in Section~\ref{sld-deter}, therefore, serve as a baseline for local linear stability, while the results from RMS computations quantify how nonsmooth friction and stochastic cutting forces reduce the practically usable parameter domain by amplifying stationary vibrations. After discarding $\tau=25000$ to remove the transients, the stationary RMS is computed as
\begin{equation}
    \mathrm{RMS} = \sqrt{\dfrac{1}{T_s}\int_{\tau_0}^{\tau_0+T_s}y(\tau)^2 d\tau}, 
\end{equation}
where $T_s$ is the duration of the stationary window. This procedure is repeated on a $250\times250$ grid in ($a_p, N$) for three different noise strength amplifications, namely $1\times$, $5\times$, and $10\times$ with the values of stochastic parameters as $\eta=0.15, \mu_{OU}=0.1, \sigma=0.2, \theta=0.7$ and the results are represented in Fig.~\ref{fig:SLD noise} with solid black line representing the deterministic stability boundary. The vertical white lines in Fig~\ref{fig:SLD noise}~(d)-(f) represent the separation between the deterministic stability boundary and RMS value corresponding to the $40\%$ of the nondimensional feed rate at $N=3600~RPM$. For representation, the difference at $40\%$ is presented as this value aligned with limits of the blue contour at $0.4$, as seen in Fig~\ref{fig:SLD noise}. This is the domain in which the stochastic limit cycle has not yet fully formed, and the dynamics remain near the fixed point. Fig~\ref{fig:SLD noise}~(a) shows the RMS value for the considered SDDE system with $1\times$ noise amplification. It is clear that stochastic vibrations are larger at higher spindle speeds. The zoomed region shown in Fig.~\ref{fig:SLD noise}~(d) around the multistable region reveals that the stochastic fluctuations move the stability boundary away from the deterministic lobes. Fig.~\ref{fig:SLD noise}~(b) and its zoomed region in Fig.~\ref{fig:SLD noise}~(e) reveal that the stable boundary is substantially displaced for $5\times$ noise amplification. Stationary RMS values for $10\times$ noise amplification and its zoomed version are represented in Figs~\ref{fig:SLD noise}~(c) and (f) respectively. It is clear that higher noise levels reduce the predictive accuracy of the deterministic boundary and also produce a broader stochastic stability envelope.}

\textcolor{black}{The separation between the deterministic stability boundary and a certain percentage of the nondimensional feed rate ($=1$), here $40\%$, at a certain spindle speed, here $N = 3600\,\mathrm{RPM}$, provides a quantitative estimate of the noise-induced erosion of stability. For the lowest noise level (1$\times$), the distance is approximately $21.6\,\mu\mathrm{m}$, indicating that the stochastic response remains close to the deterministic prediction. As the noise intensity is increased to 5$\times$ and 10$\times$, this separation grows to $52.3\,\mu\mathrm{m}$ and $83.3\,\mu\mathrm{m}$, respectively. This monotonic increase demonstrates that stronger stochastic forcing shifts the effective stability limit to shallower cut depths, causing chatter-like vibration levels to appear well within the region nominally stable according to the deterministic lobe. These results suggest that the deterministic lobes capture the nominal stability limit, stochastic fluctuations introduce a clear shift of the effective safe region of operation.} 

\begin{figure}[!hbt]
    \centering
    \includegraphics[width=0.65\linewidth]{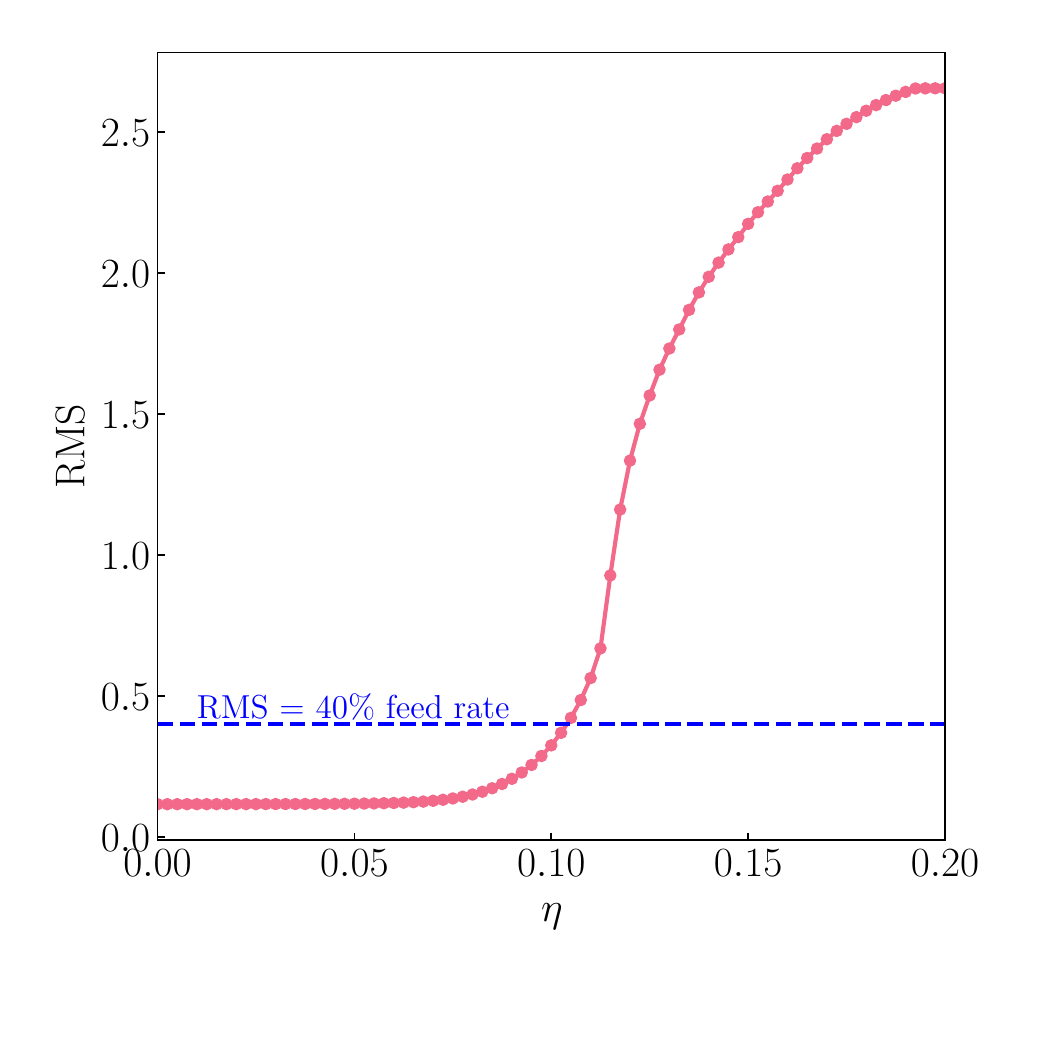}
    \caption{\textcolor{black}{Stationary RMS value of the nondimensional tool displacement for various values of the noise intensity at $a_p=0.54$ and $N=3600$.}}
    \label{fig:etaVSrms}
\end{figure}
\textcolor{black}{Stationary RMS value of the nondimensional tool displacement for various values of the noise intensity for $a_p=0.54$ and $N=3600$ is represented in Fig.~\ref{fig:etaVSrms}. The analysis is carried out by varying the noise intensity $\eta$ from 0 to 0.2, while keeping the other stochastic parameters fixed at $\mu_{OU}=0.1, \sigma=0.2, \theta=0.7$. The dotted blue line in Fig.~\ref{fig:etaVSrms} denotes the RMS level corresponding to $40\%$ of the dimensionless feed rate, matching the blue contour at 0.4 in Fig.~\ref{fig:SLD noise}. From Fig.~\ref{fig:etaVSrms}, the RMS amplitude remains approximately constant at 0.1 for $\eta$ values between 0 and 0.07, after which it begins to rise. The curve crosses the $40\%$ feed-rate threshold at $\eta=0.104$, indicating that the transition to higher vibration levels occurs at this point. For the subsequent analysis, $\eta$ is fixed at 0.15 to ensure that the influence of noise is clearly present.}

\par A stochastic attractor can be defined as a set $A_S \subset \mathbb{R}^n$ such that once the system is in the bounded region $A_S$, it remains in $A_S$ for a finite time $t$ under consideration despite the influence of stochastic fluctuations. For any initial state $\mathbf{X}_0$ in the basin of the attractor, $P(\mathbf{X}_t \in A_S) \rightarrow 1$ as $t\rightarrow \infty$ \cite{crauel1997random}. The systems with stochastic excitations have been studied under the purview of two distinct classes of bifurcations, namely, P-bifurcation and D-bifurcation\cite{arnold1996toward,crauel1997random,kumar2017bifurcation}.
\begin{figure*}[!hbt]
     \centering
          \includegraphics[width=\textwidth]{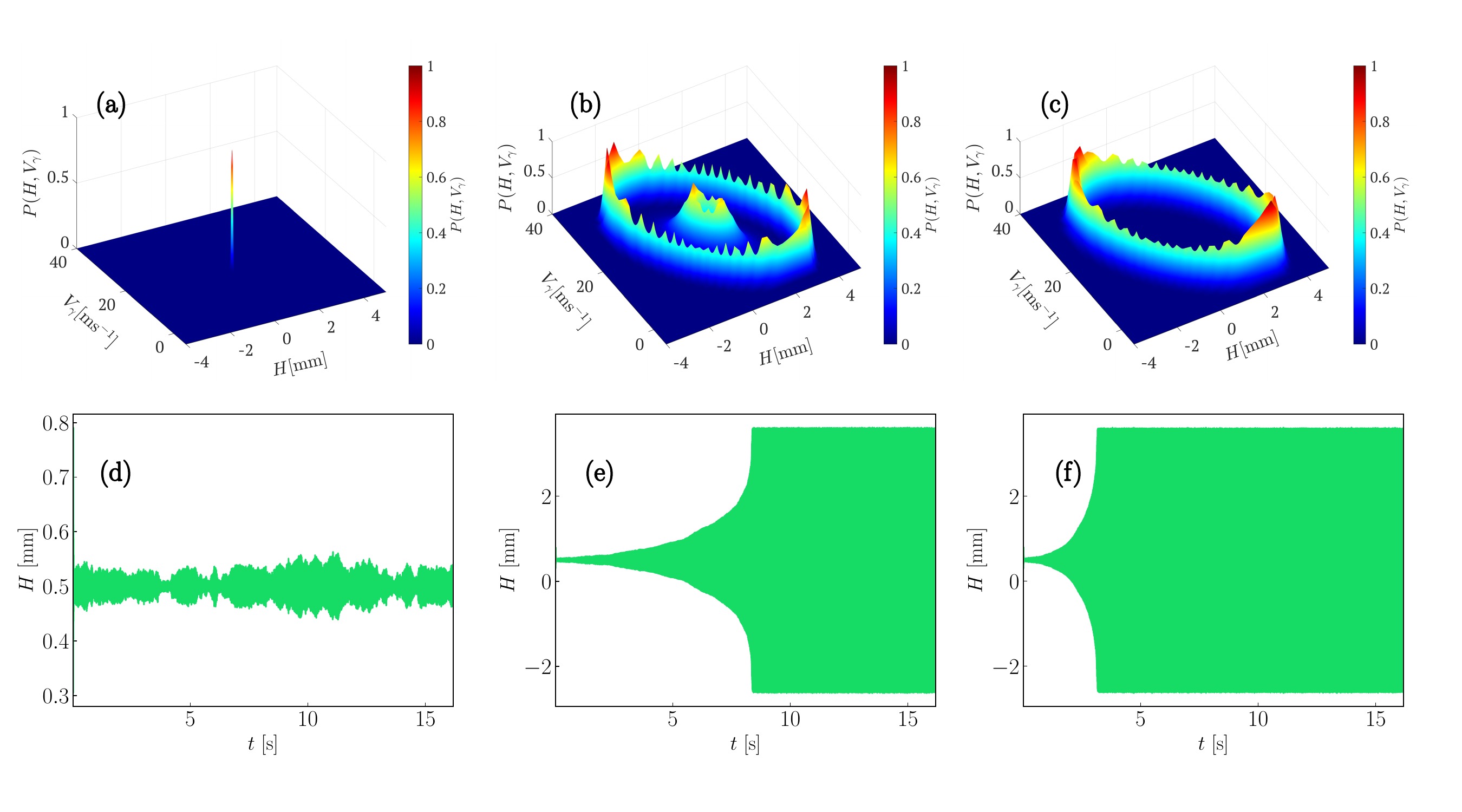}
         \caption{Stochastic P-bifurcation of the metal cutting system for the parameter values of $\eta=0.15, \mu_{OU}=0.1, \sigma=0.2, \theta=0.7$. Figures (a)-(c) show the probability density values for depth of cut values of $a_p = 0.52$ $mm$, $0.54$ $mm$, $0.55$ $mm$. (d)-(f) show the corresponding chip thickness over time.}
         \label{fig:p-bifur1}
\end{figure*}

\subsection{\textrm{P-bifurcation}}
\begin{figure*}[!hbt]
     \centering
          \includegraphics[width=\textwidth]{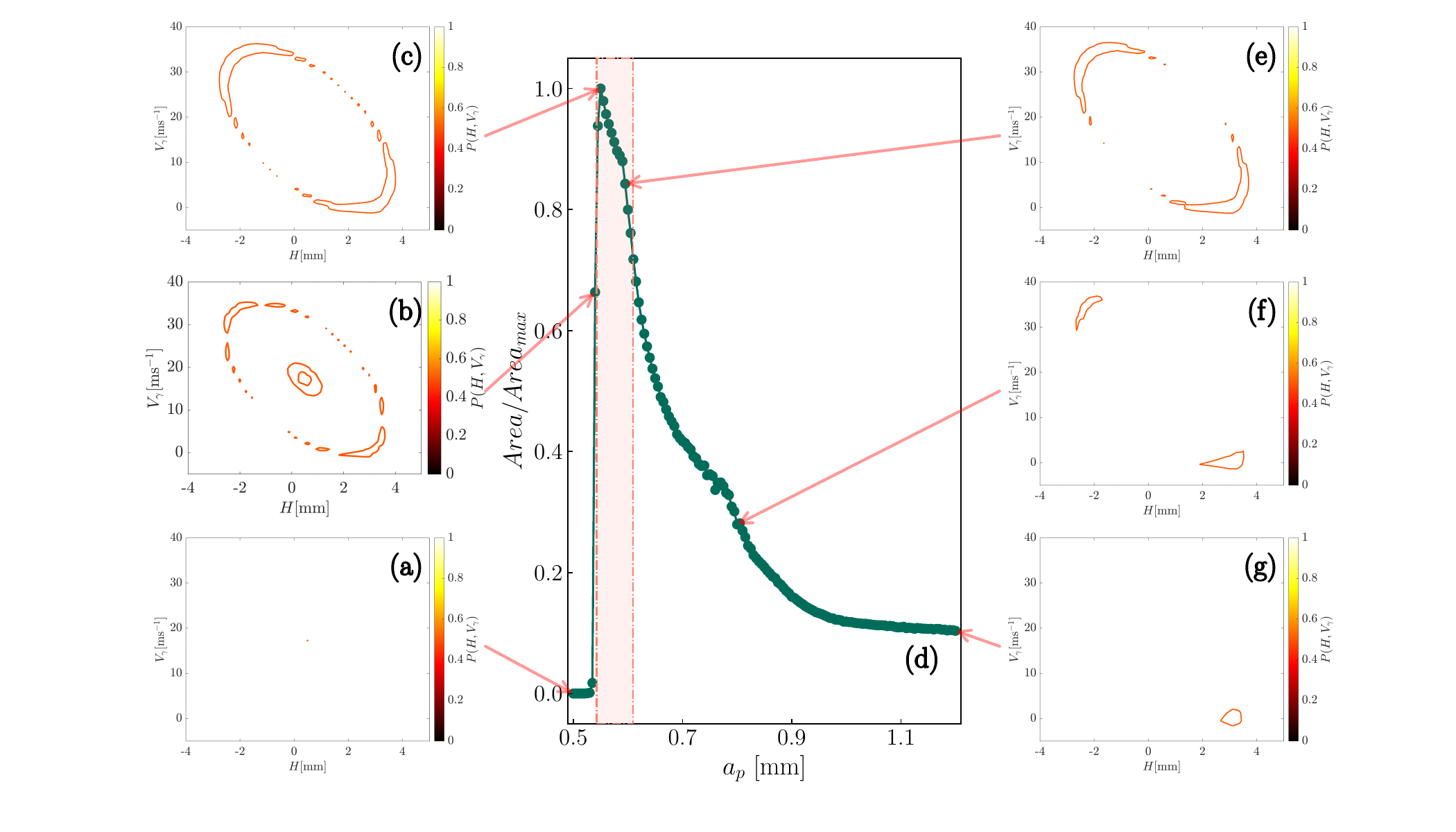}
         \caption{Figure (d) representing the area of the contours at j-pdf = 0.5. The shaded region denotes the multi-stable region. The left and right panels show the contour plots of probability equal to 0.5 for specific depth of cut $(a_p)$ values, namely, (a)~0.5, (b)~0.54, (c)~0.55, (e)~0.595, (f)~0.8, and (g)~1.2 $mm$.}
         \label{fig:p-bifur2}
\end{figure*}

\noindent Phenomenological or \textit{P-bifurcations} arise when changes in a system parameter cause abrupt topological shifts in the probability distribution governing its long-term behavior \cite{kumar2017bifurcation, arnold1996toward}. The joint probability density (\textit{j-pdf}) represents a statistical measure of how much time a typical system trajectory spends within a given volume element in the phase space. It also indicates the spatial extent of a stochastic attractor \cite{kumar2017bifurcation}. Fig.~\ref{fig:p-bifur1} depicts the occurrence of the \textit{P-bifurcation} in the considered system where the stochastic parameters are taken as $\eta=0.15, \mu_{OU}=0.1, \sigma=0.2, \theta=0.7$. Figs.~\ref{fig:p-bifur1}~(a)-(c) correspond to the \textit{j-pdf }for three different bifurcation parameter values $a_p = 0.52\;mm$, $a_p=0.54\;mm$ and $a_p=0.55\;mm$ respectively. It can be observed that there is a peak in Fig.~\ref{fig:p-bifur1}~(a), which indicates the fixed point attractor. Fig.~\ref{fig:p-bifur1}~(b) shows a crater-like shape with a peak inside, indicating the occurrence of a transition of the dynamical state from one underlying deterministic attractor to another. The stability of the system switches from the non-chatter to the chatter region under the influence of noise. Further, the inside peak disappears in Fig.~\ref{fig:p-bifur1}~(c) with the further increase of $a_p$. The corresponding changes in the dynamics are shown in Figs.~\ref{fig:p-bifur1}~(d)-(f), which present the instantaneous chip thickness. In particular, it is evident from Fig.~\ref{fig:p-bifur1}~(e) that the attractors corresponding to both the non-chatter and the chatter cutting appear to have non-zero probabilities of occurrence when the dynamics is observed for a sufficiently long time. In this analysis, time histories were simulated under the assumption of ergodicity. This assumption implies that time-averaged statistics obtained from a single, sufficiently long trajectory are equivalent to ensemble-averaged quantities across multiple realizations \cite{arnold1996toward}. The simulations are carried out for $\tau=5.5\times10^4$. The \textit{j-pdf }values are numerically computed using the kernel smoothing function estimate \textit{ksdensity} in MATLAB. To visualize the P-bifurcation quantitatively, the area of the contours of the \textit{j-pdf} at a specific value of probability $P=0.5$ is computed using a MATLAB routine \textit{Contour2Area} \cite{Sundqvist2025Contour2Area} and the results are shown in Fig.\ref{fig:p-bifur2}. In this calculation, the values of $a_p$ are set to 0.5-1.2 $mm$ with a step size of 0.005. Fig.~\ref{fig:p-bifur2}~(d) displays the normalized area corresponding to contour levels at a \textit{j-pdf }value of 0.5. Sub-figures (a)-(c) and (e)-(g) illustrate how the contour shapes (\textit{j-pdf}=0.5) evolve across various values of $a_p$, specifically, (a) 0.5 $mm$, (b) 0.54 $mm$, (c) 0.55 $mm$, (e) 0.595 $mm$, (f) 0.8 $mm$, and (g) 1.2 $mm$. The morphological changes observed in these contours are reflected quantitatively in the area trend shown in Fig.~\ref{fig:p-bifur2}~(d). It is observed that there is a sudden jump in the area exactly at the beginning of the multi-stable region $a_p=0.54\; mm$, which coincides with the topological changes shown in Fig.~\ref{fig:p-bifur1}, indicating that a P-bifurcation has taken place.

\subsection{\textrm{D-bifurcation}}
\par
\begin{figure}[!h]
     \centering
          \includegraphics[width=1.0\linewidth]{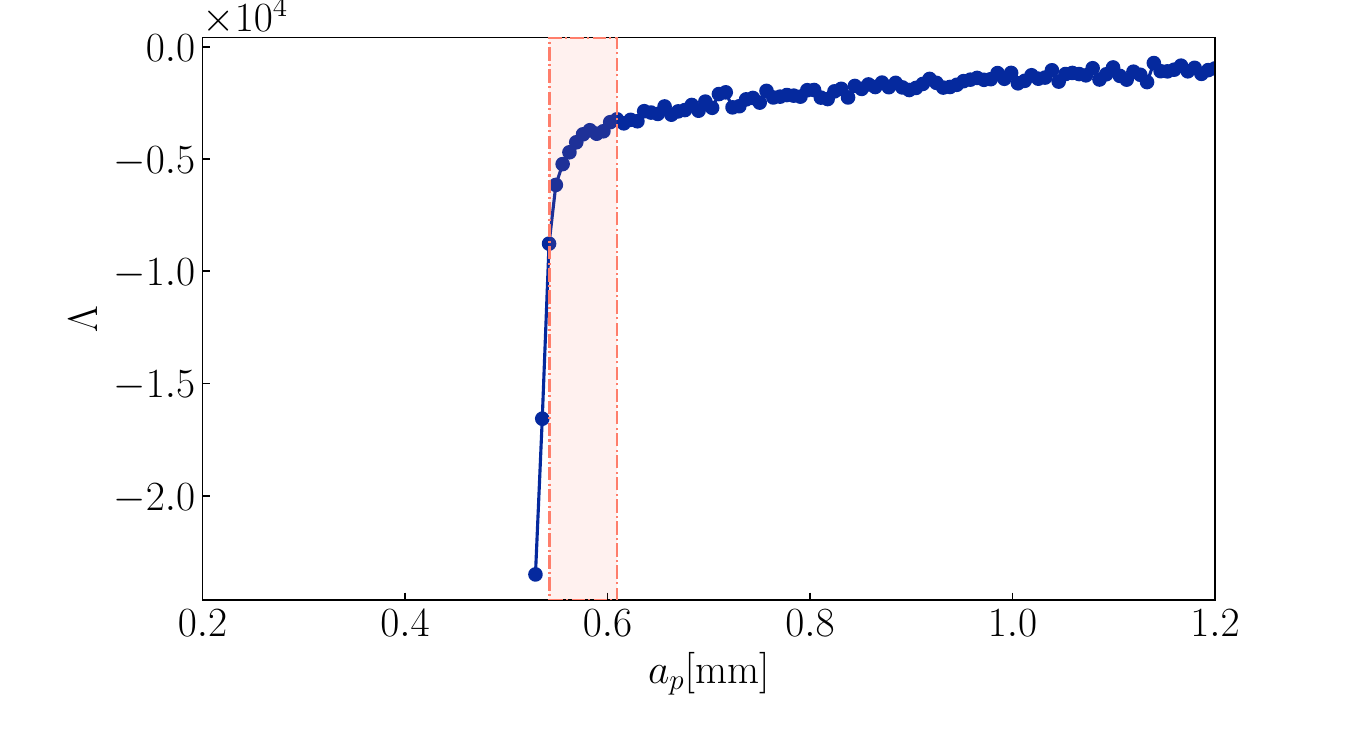}
         \caption{Estimation of Lyapunov exponent with $a_p$ as the bifurcation parameter.}
         \label{fig:d-bifur}
\end{figure}

 Dynamical or \textit{D-bifurcations} refer to sudden topological changes in the underlying phase-space dynamics \cite{kumar2017bifurcation}. Unlike \textit{P-bifurcations}, which characterize qualitative changes in the shape of \textit{j-pdf}, \textit{D-bifurcations} reflect shifts in the dynamical stability of stochastic attractors. Stability is assessed by calculating Lyapunov exponents (LE), which quantify the average exponential rate of divergence or convergence of nearby trajectories. In stochastic systems with continuous perturbations, a negative LE indicates that perturbations remain bounded and the system tends toward a stochastic attractor, even though individual fluctuations persist due to the added noise. For the \textit{D-bifurcation} analysis following approximation for LE has been used \cite{ddechaosreview2019wernecke}.

\begin{equation}
    \Lambda = \lim_{\tau\rightarrow\infty}\lim_{\delta\rightarrow 0} 
    \mathbb{E}\left(\dfrac{1}{\tau}\log\left(\dfrac{\left|y(\tau)-y_{pert}(\tau)\right|^2}{\delta}\right)\right),
\end{equation}
where $y_{pert}(\tau)$ is the perturbed trajectory with small perturbation $\delta$ with respect to the original trajectory $y(\tau)$ and $\mathbb{E}$ denotes the expectation operator. Fig.~\ref{fig:d-bifur} visualizes the LE values $\Lambda$ as a function of the depth of cut $a_p$. 
From Fig.~\ref{fig:d-bifur}, one can observe that LE has negative values over all the values of $a_p$. This is due to the stable nature of both attractors. However, there are no data points in Fig.~\ref{fig:d-bifur} for $a_p$ ranging from $0.2$ to $0.53\;mm$, as the corresponding LE values approached $-\infty$, indicating that the perturbations were too small to produce divergence. The first point in Fig.~\ref{fig:d-bifur} is around $a_p=0.53\;mm$ where the transition in the dynamics occurred and has a gradually increasing trend at the start of the multi-stable region and settled around $-1.2\times10^4$. The D-bifurcation occurs at $a_p=0.53\;mm$, whereas from Fig.~\ref{fig:p-bifur1}~(b) it is evident that the peak structure still exists inside the crater-like structure and it disappears at $a_p=0.55\;mm$. It is important to mention here that the \textit{D-bifurcation} occurs at a different value of the bifurcation parameter than the \textit{P-bifurcation}\cite{kumar2017bifurcation}.
\begin{figure*}[!hbt]
     \centering
          \includegraphics[width=\textwidth]{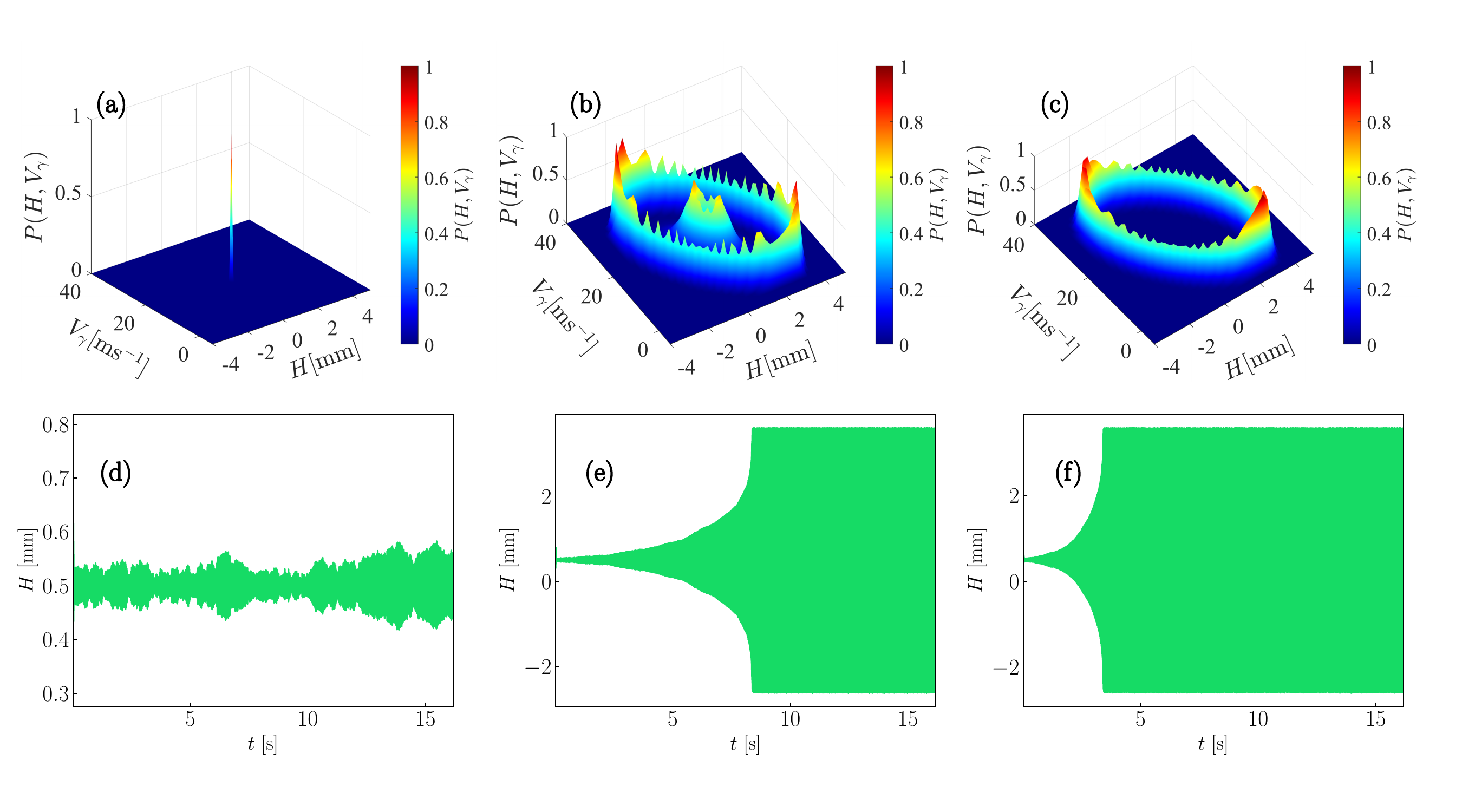}
         \caption{\textcolor{black}{Stochastic P-bifurcation of the metal cutting system for the parameter values of $\eta=0.15, \mu_{OU}=0.1, \sigma=0.2, \theta=0.7$. Figures (a)-(c) show the probability density values for depth of cut values of $N = 3610$ $RPM$, $3600$ $RPM$, $3590$ $RPM$. (d)-(f) show the corresponding chip thickness over time.}}
         \label{fig:p-bifur1-vary-N}
\end{figure*}
\subsection{\textcolor{black}{Influence of Time-delay}}
\begin{figure*}[!hbt]
     \centering
          \includegraphics[width=\textwidth]{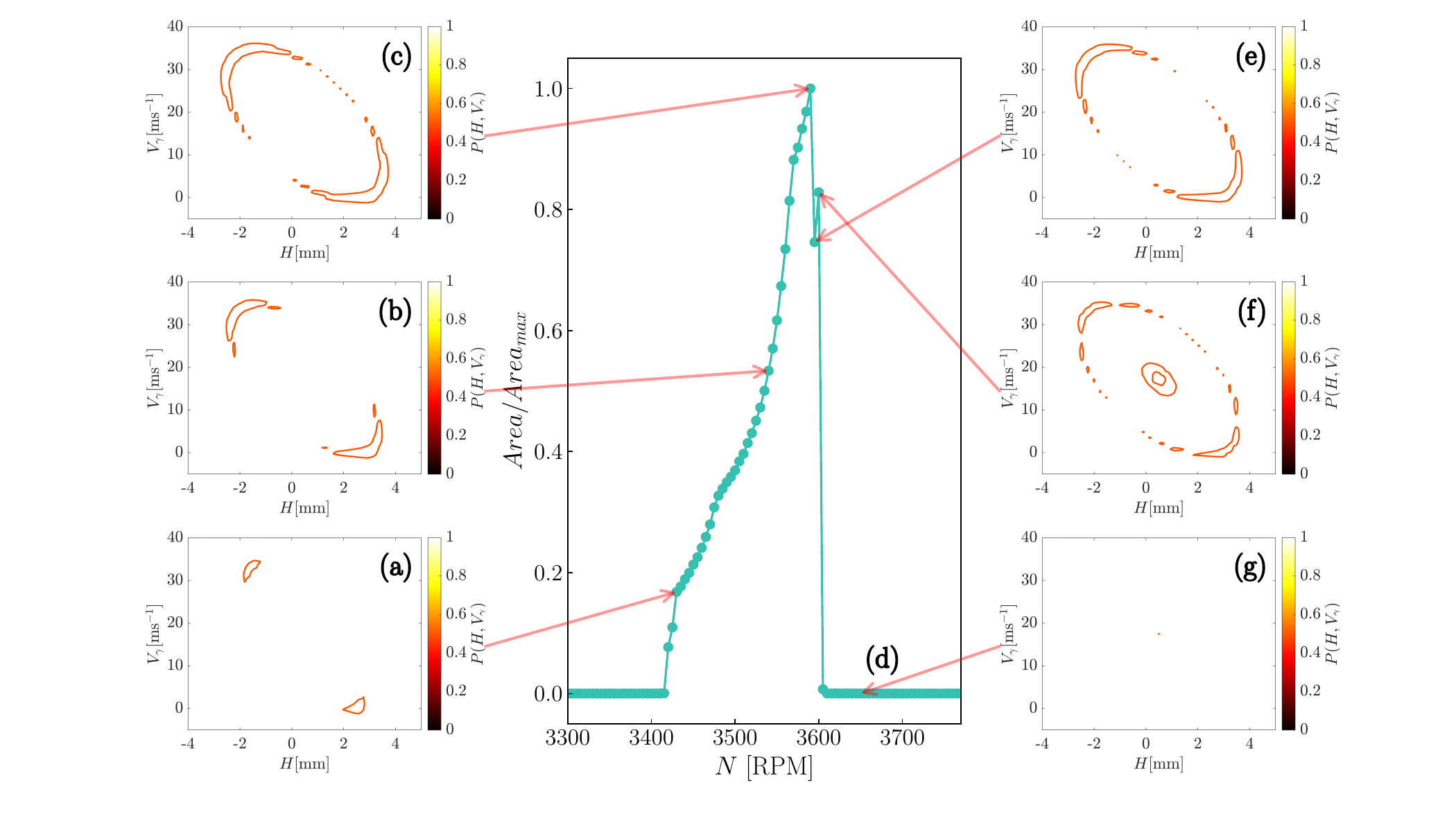}
         \caption{\textcolor{black}{Figure (d) representing the area of the contours at j-pdf = 0.5. The left and right panels show the contour plots of probability equal to 0.5 for specific spindle speed $N$ values, namely, (a)~3430, (b)~3540, (c)~3590, (e)~3595, (f)~3600, and (g)~3650 $RPM$.}}
         \label{fig:p-bifur2-vary-N}
\end{figure*}

\begin{figure}[!h]
     \centering
          \includegraphics[width=1.0\linewidth]{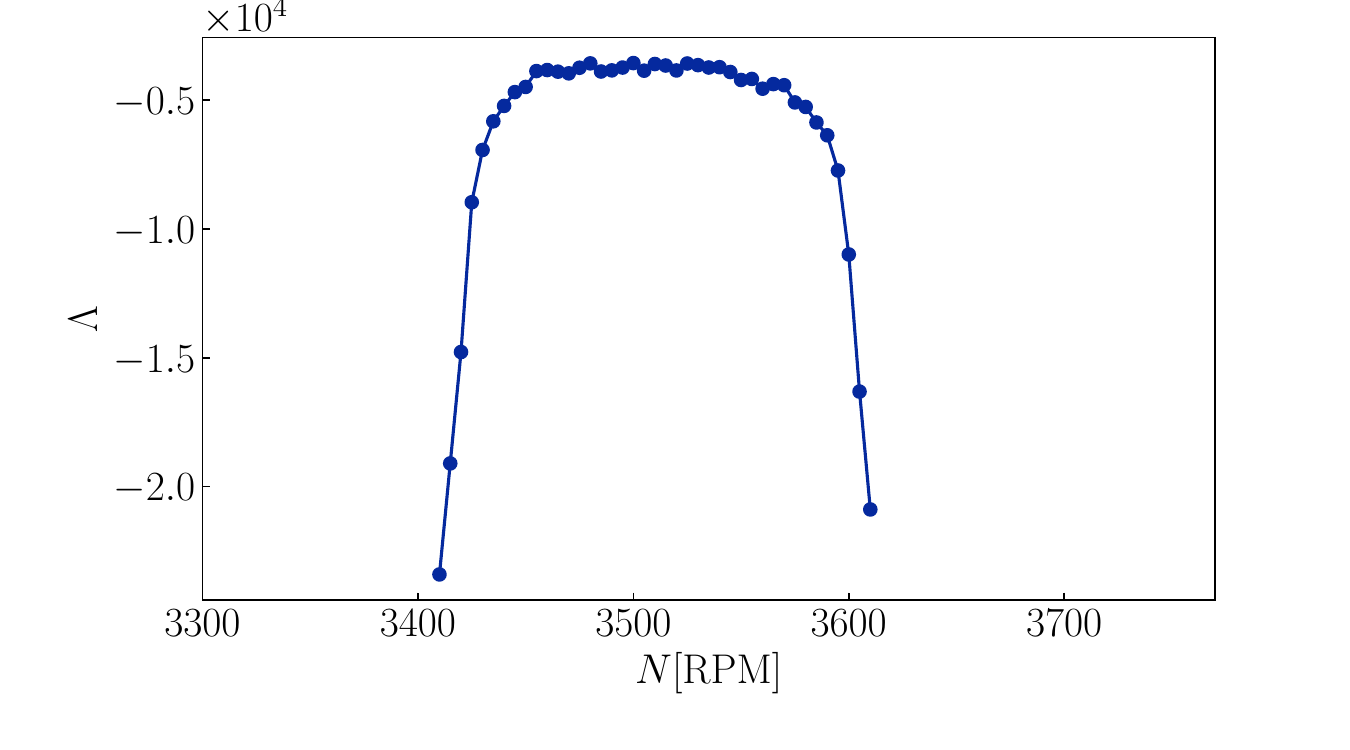}
         \caption{\textcolor{black}{Estimation of Lyapunov exponent with $N$ as the bifurcation parameter.}}
         \label{fig:d-bifur-vary-N}
\end{figure}

\textcolor{black}{To investigate the influence of time-delay on stochastic bifurcations, the same analyses are carried out for varying time-delay parameter. In these analyses, the depth of cut $a_p$ was fixed at a constant value of $0.54~mm$ while the non-dimensional time-delay parameter $\tau_w$ was varied from $54.08987$ to $61.7936$ by adjusting the dimensional spindle speed ($N$) within the range of $3300$ to $3770~RPM$. Results are presented as a function of $N$ to align with other stability analyses, such as SLDs, which use $a_p$ and $N$. All other parameter values are considered as in the previous section. Fig.~\ref{fig:p-bifur1-vary-N} represents the occurrence of P-bifurcation for various values of $N$. Figs.~\ref{fig:p-bifur1-vary-N} (a)-(c) correspond to $N=3610~RPM$, $N=3600~RPM$, and $3590~RPM$ respectively. The results observed here are similar to those of the P-bifurcation results with varying $a_p$. The quantitative analysis of P-bifurcation with the area calculation of the contours of the \textit{j-pdf} at $P=0.5$ is sketched in Fig.~\ref{fig:p-bifur2-vary-N}. The normalized area corresponding to contour levels at  \textit{j-pdf=0.5} for $N$ varying from $3395$ to $3770~RPM$ given in Fig.~\ref{fig:p-bifur2-vary-N}~(d). Sub-figures in left and right panels of Fig.~\ref{fig:p-bifur2-vary-N} display how the contour shapes change over the change of the $N$ values, particularly, (a)~3430, (b)~3540, (c)~3590, (e)~3595, (f)~3600, and (g)~3650. The \textit{D-bifurcation} analysis was also carried out with the same settings for varying spindle speed values. The outcomes of LE computations are illustrated in Fig.~\ref{fig:d-bifur-vary-N}. From Fig.~\ref{fig:d-bifur-vary-N}, one can observe that there are no data points for $N$ ranging from 3300 to 3410 and from 3615 to 3770, indicating that the LE values have approached $-\infty$. The first transition in D-bifurcation occurred at  $N=3410~RPM$, whereas the first transition in P-bifurcation occurred at $N=3425~RPM$. Also, the second transition happened at $N=3615~RPM$ and $N=3600~RPM$ for D- and P-bifurcations, respectively. A similar trend was observed in the previous analysis, with $a_p$ as the bifurcation parameter. These results are summarized in table~\ref{tab:p-d-bifur}.}

\begin{table}[!hbt]
    \centering
    \begin{tabular}{|c|c|c|}
        \hline
         Bifurcation Parameter & P-bifurcation & D-bifurcation\\ \hline
         $a_p~[mm]$ & 0.54 & 0.53\\
         \hline
         \multirow{2}{4em}{$N~[RPM]$} & 3425 & 3410\\
               &3600 & 3615\\ \hline
    \end{tabular}
    \caption{P- and D-bifurcations occurring points}
    \label{tab:p-d-bifur}
\end{table}

\section{Stability of Stochastic Attractors}\label{sec5}
\noindent In the previous section, the computations were performed under the assumption that, for a given bifurcation parameter, the initial history function remained identical across all trials. However, experimental studies show that the initial surface profiles of the workpiece have a significant effect on the stability of the dynamics of cutting tools \cite{benardos2003predicting}. This necessitates studying uncertainty in initial surface profiles, which is done using the method of basin stability analysis. Monte Carlo simulations have been carried out with continuously present perturbations and variations in initial history functions. The regimes to be operated into avoiding chatter, particularly in the multi-stable region, have been demarcated. Furthermore, entropy metrics are carried out to characterize transitions in system dynamics by analyzing the statistical structure of the underlying time histories.

\subsection{Basin Stability Analysis}\label{sec-basin}
\noindent Basin stability for a system without delay is normally computed by considering the percentage of initial conditions settling into each attractor in the dynamical system\cite{belardinelli2017improving}. However, calculating basins of attraction in the same manner for a DDE needs reconsideration, as such systems have infinite initial conditions. The computation of basin stability analysis for the stochastic case is adapted from the deterministic basin stability computation as given in Refs .~\cite{turning-yan2021safety, yan2017basins}. The expressions for the initial history function are considered as

\begin{subequations}
    \begin{equation}
        y_1(\tau) = s(\theta_{bs}(\tau))-p_u(\theta_{bs}(\tau)),
    \end{equation}\label{eq:bs_fun}
    where
    \begin{equation}\label{eq:bs_fun2}
        p_u(\theta_{bs}(\tau)) = 1-\dfrac{\theta_{bs}(\tau)}{2\pi},
    \end{equation}
    \begin{equation}\label{eq:bs_fun3}
        \theta_{bs}(\tau) = 2\pi+\tau\Omega,\;\;  \tau \in [-\tau_w, 0),
    \end{equation}
    and the $s(\theta_{bs}(\tau))$ represented by the following Fourier series
    
    \begin{equation}\label{eq:bs_fun4}
        \begin{aligned}[b]
            s(\theta_{bs}(\tau)) =&\sum_{i=1}^{N_{bs}}a_i (\sin(i\Omega\tau+i\phi_{bs}) - \sin(i\phi_{bs})) \\
            &+ \sum_{i=1}^{N_{bs}}b_i (\cos(i\Omega\tau+i\phi_{bs}) - \cos(i\phi_{bs})),
        \end{aligned}
    \end{equation}
\end{subequations}
\begin{figure}[!hbt]
     \centering
          \includegraphics[width=0.6\linewidth]{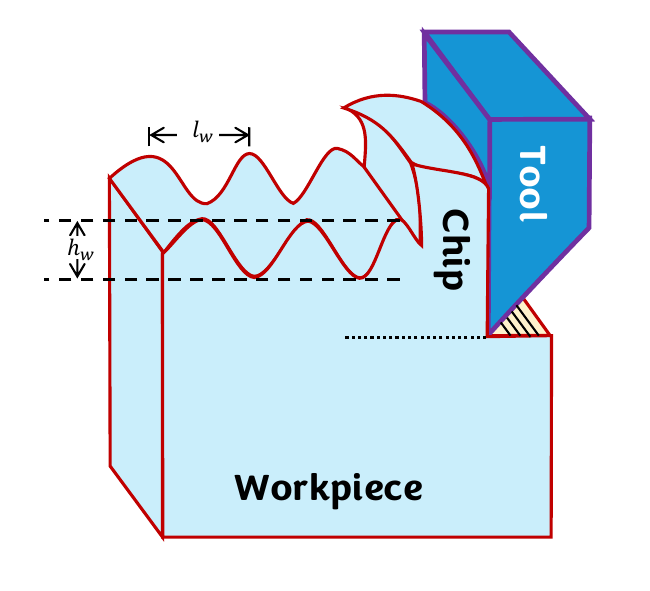}
         \caption{Schematic figure showing initial waviness and height}
         \label{fig:basin_sche}
\end{figure}
\noindent where the coefficients $a_i$ and $b_i$ specify the initial waviness length and height of the workpiece surface. Fig.~\ref{fig:basin_sche} shows the schematic diagram of waviness height ($h_w$) and length ($l_w$). $\phi_{bs}\in[0,2\pi)$ represents the tool-workpiece engagement geometry at the onset of cutting. To set the limits on waviness length and height, the following constraint was used in the initial function when calculating basin stability.
\begin{equation}
\label{Eq-constraint-alpha}
    \sum_{i=1}^{N_{bs}}a_i^2+b_i^2 \leq \alpha^2.
\end{equation}

\begin{figure*}[!hbt]
     \centering
          \includegraphics[width=0.7\linewidth]{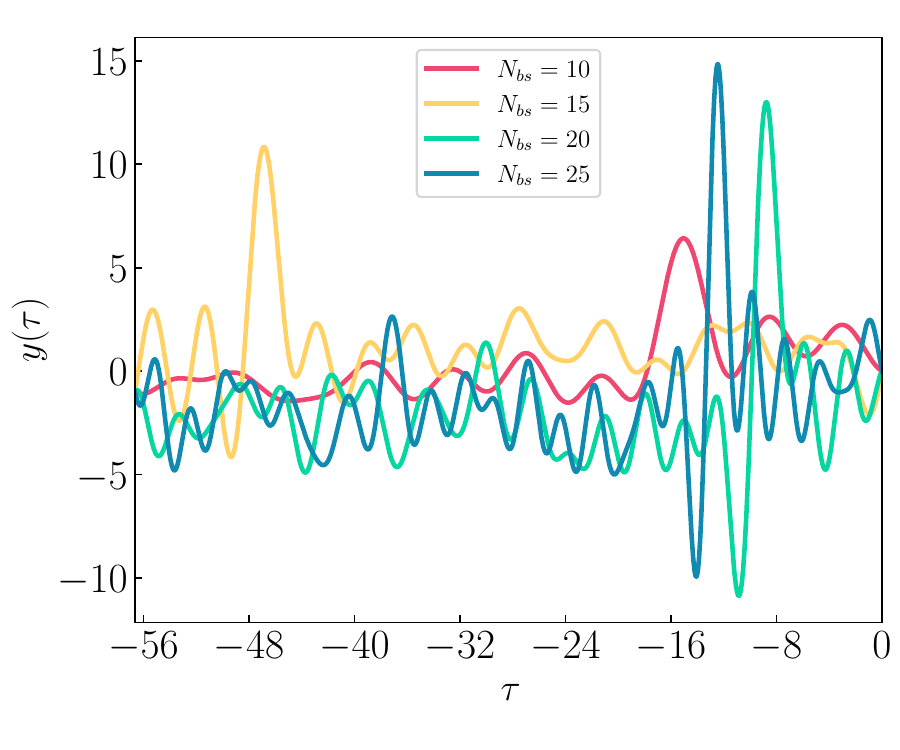}
         \caption{\textcolor{black}{Initial profile for various value of $N_{bs}$.}}
         \label{fig:compar_N_bs}
\end{figure*}

\textcolor{black}{The $\alpha$ given in Eq.~\eqref{Eq-constraint-alpha} controls the overall initial profile waviness on the workpiece. In other words, the sum of squared amplitudes of $a_i$ and $b_i$ in Eq.~\eqref{eq:bs_fun4} cannot exceed $\alpha^2$, where $a_i$ and $b_i$ are the amplitudes of the sine and cosine components of the $i^{th}$ waviness mode. For a prescribed number of basis functions $N_{bs}$, two coefficient vectors $a \in \mathbb{R}^n$ and $b \in \mathbb{R}^n$ are first drawn with independent entries from a uniform distribution on $[0,1]$. To ensure that the overall amplitude of the resulting Fourier series remains bounded, a constraint given in Eq.\eqref{Eq-constraint-alpha} is then enforced on the coefficients. If the coefficient vectors satisfy Eq.\eqref{Eq-constraint-alpha}, the coefficients are left unchanged. Otherwise, all coefficients are uniformly rescaled by the positive scaling factor $\left(\sqrt{\dfrac{\alpha^2}{\sum_{i=1}^{N_{bs}}a_i^2+b_i^2}}\right)$ applied to every entry of both vectors.}
\begin{figure*}[!hbt]
     \centering
          \includegraphics[width=0.7\linewidth]{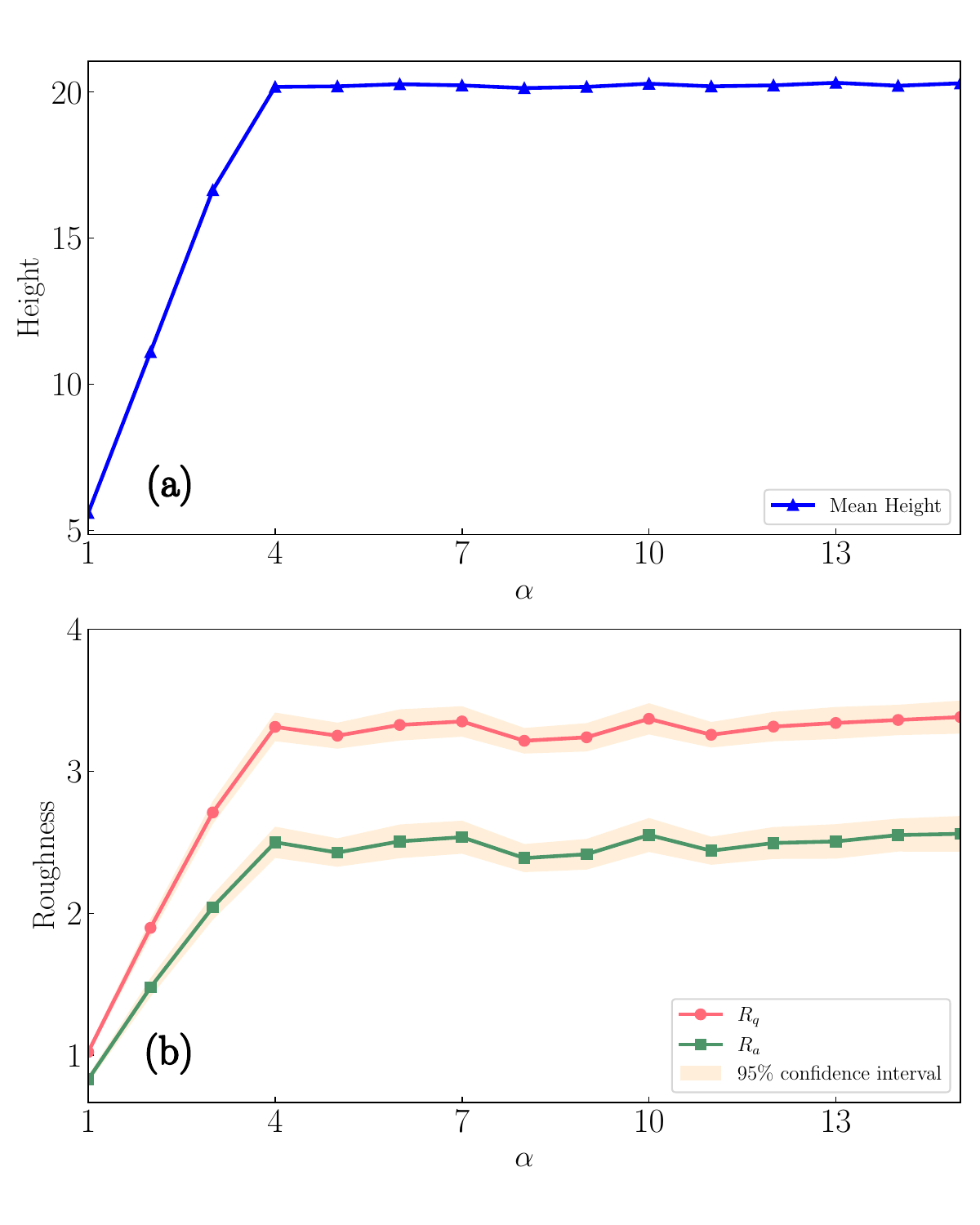}
         \caption{\textcolor{black}{(a)~Comparison of initial profile height and (b)~Values of $R_a$ and $R_q$ for ranges of $\alpha$ values.}}
         \label{fig:height_ra_rq}
\end{figure*}
\textcolor{black}{To show the impact of the constraint parameter $\alpha$ and the number of basis functions $N_{bs}$ on the initial surface waviness on the workpiece, calculation of average roughness $R_a $, root mean square roughness $R_q $, and height of the initial profile are calculated as functions of various values of $\alpha$. The surface roughness measures are computed numerically using a Monte Carlo-based stochastic simulation procedure. First, the regenerative delay interval is discretized over the time domain $( \tau \in [-\tau_w,0] )$ using a sufficiently small time step ($10^{-3}$). For each value of $\alpha$, a number of stochastic realizations ($M_{sto}=1000$) are generated to capture the variability of the surface profile. For each realization, a random phase angle and a set of random Fourier coefficients are generated, and the initial trajectory is obtained for $\tau \in [-\tau_w,0]$ from Eq.~\eqref{eq:bs_fun}. From this trajectory, the arithmetic average roughness $R_a $ and the root mean square roughness $R_q $ are evaluated as
\begin{equation}
    R_a=\frac{1}{M_{sto}}\sum_{i=1}^{M_{sto}}|y_{1_i}(\tau)|,
\end{equation}
and
\begin{equation}
    R_q=\sqrt{\frac{1}{M_{sto}}\sum_{i=1}^{M_{sto}}y^2_{1_i}(\tau)},\qquad\tau \in [-\tau_w,0],
\end{equation}
Also, the peak-to-valley height of the profile is determined by the difference between its maximum and minimum values along the trajectory. This process is repeated for all Monte Carlo samples, and the ensemble mean and standard deviation of the roughness measures are subsequently computed. The statistical uncertainty is quantified using a $95\%$ confidence interval. Finally, the mean values, together with their confidence intervals, are plotted as functions of the constraint parameter $\alpha$. The results are represented in Fig.~\ref{fig:height_ra_rq}. It is clear from Fig.~\ref{fig:height_ra_rq}~(a) that the peak-to-valley height of the initial profile is smaller for small values of $\alpha$ and settles around 20 for $\alpha$ values higher than 4. A similar trend was observed in Fig.~\ref{fig:height_ra_rq}~(b) for roughness calculation as well. The calculated values have no units because they are computed from the history function of the non-dimensional equation. To relate to physical machining conditions, the simulated nondimensional roughness levels can be qualitatively associated with the initial surface roughness of the workpiece used in turning processes. Here $\alpha$ serves as a calibrated roughness-control parameter that orders the workpiece profiles from smooth to rough. The influence of the number of basis functions $N_{bs}$ on the initial profile is also shown in Fig.~\ref{fig:compar_N_bs}.}

Also, to restrict the strength of initial tool vibrations, another constraint is considered as follows
\begin{subequations}
  \begin{equation}
    y_1(0)\in [-\beta, \beta],
  \end{equation}  
  \begin{equation}
    y_2(0)\in [-\beta, \beta],
  \end{equation} 
\end{subequations}
\textcolor{black}{where $\beta$ denotes the allowable initial vibration level of the cutting tool prior to the process reaching its asymptotic regime. A smaller $\beta$ indicates that the process is more strictly controlled with the tool start-up condition, while a larger $\beta$ allows greater initial perturbations, potentially increasing the likelihood of transition to chatter. As the displacement is nondimensionalized by the value of nominal chip thickness (feed-rate) $H_D$ as $y_1(\tau)=Y(t)/H_D$ (Eq.~\eqref{Eq-nondimY}), the corresponding dimensional bound on the initial tool displacement is $\beta\times H_D~[mm]$. Therefore, $\beta$ serves as a constraint on the admissible initial state space and is employed to assess the sensitivity of the cutting dynamics to the initial tool displacement.}

\begin{figure}[!hbt]
     \centering
          \includegraphics[width=0.8\linewidth]{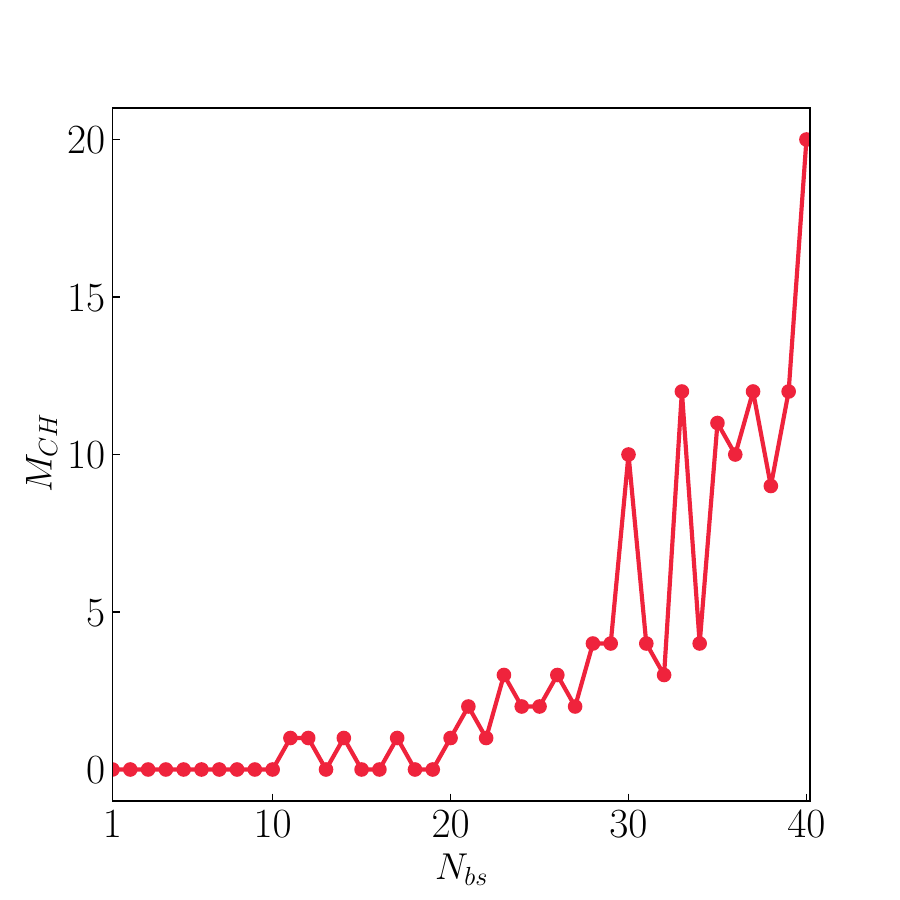}
         \caption{Basin stability of chatter cutting is estimated and displayed as functions of the number of basis functions $N_{bs}$. The constraint parameters are $\alpha = \beta = 6$. The $y$-axis shows the number of samples that lead to chatter cutting.}
         \label{fig:basin_single}
\end{figure}

The basin stability in the multi-stable region can be computed by considering the initial history function \eqref{eq:bs_fun}, based on Monte Carlo simulations with $M_{bs}$ groups of coefficients, $\{a_1^{(j)},...,a_n^{(j)}, b_1^{(j)},...,b_n^{(j)}, y_1(0)^{(j)}, y_2(0)^{(j)}, \phi_{bs}^{(j)}\}_{j=1}^{M_{bs}}$. The stochastic version of the considered model \eqref{eq-nondim-stoch} is solved using the initial function \eqref{eq:bs_fun} approximation for the following parameter values: stochastic process parameters $\eta=0.15, \mu_{OU}=0.1, \sigma=0.2, \theta=0.7$, the system parameters $a_p=0.54\;~mm$, $N=3600$~RPM, and the constraints are fixed as $\alpha=\beta=6$. In this study, a total of $M_{bs}=1000$ Monte Carlo simulations were performed, each run over a non-dimensional time ($\tau$) of $1.1\times 10^4$. Fig.~\ref{fig:basin_single} shows the number of samples leading to the sticking limit cycle attractor $M_{CH}$ as a function of the number of basis functions $N_{bs}$.
\begin{figure*}[!hbt]
     \centering
          \includegraphics[width=1.0\linewidth]{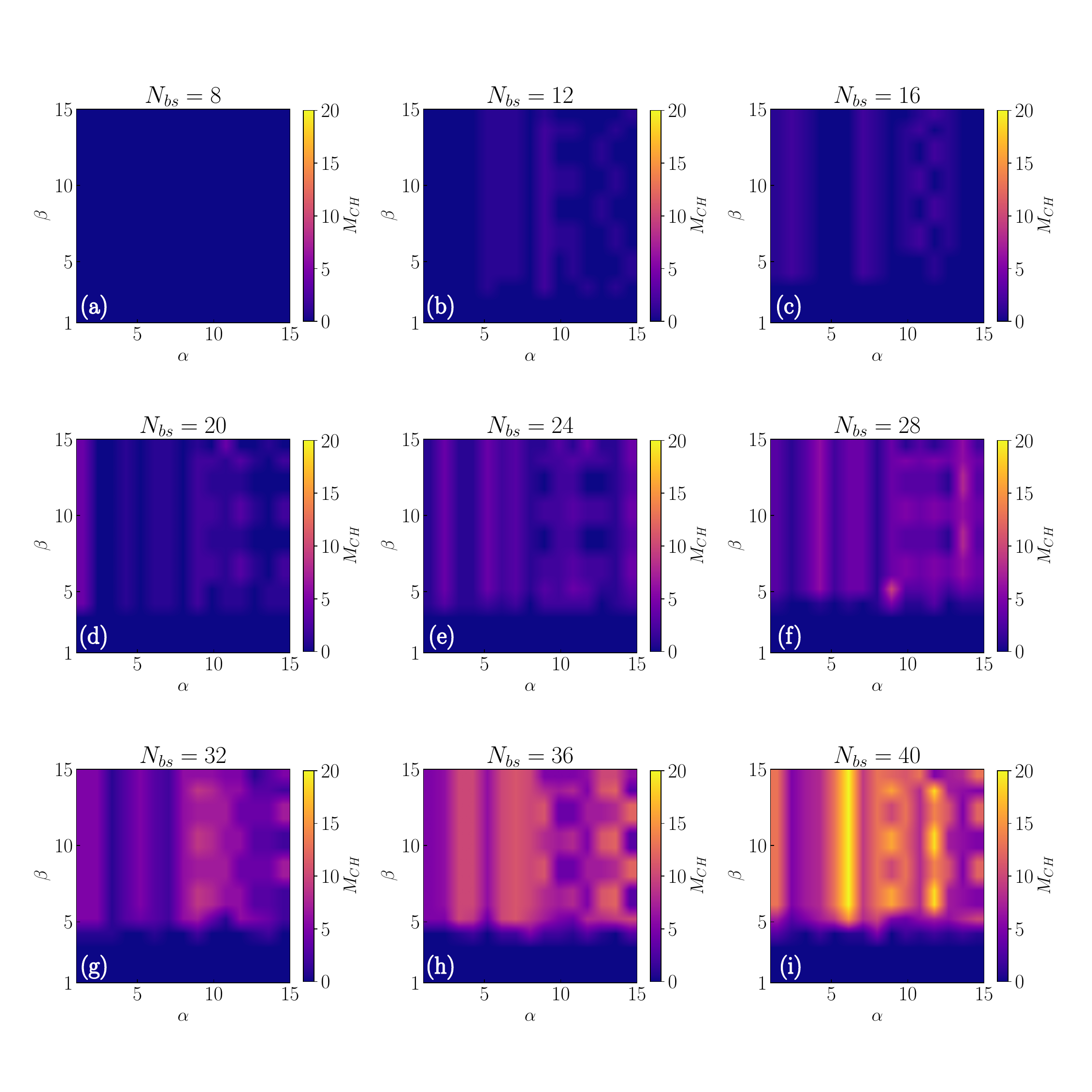}
         \caption{Influence of $\alpha$ and $\beta$ on basin stability of chatter cutting for various values of the number of basis functions $N_{bs}$.}
         \label{fig:basin_full}
\end{figure*}
It is clear that increasing the number of basis functions, $N_{bs}$, results in a more uneven workpiece surface, which, in turn, affects the tool's stability. When $N_{bs}$ exceeds 20, $M_{CH}$ begins to increase, reaching its maximum value of 20 at $N_{bs}=40$. A higher initial surface roughness, with $M_{CH} \geq 10$, holds a non-zero probability of driving the system to a state of chatter under noisy fluctuations. The same analysis has been carried out by tweaking the constraint parameters, depicting varied heights and widths of irregularities in the workpiece, initial tool displacement, and the results are plotted in $\{\alpha,\beta\}$ space for various values of $N_{bs}$ in Fig.~\ref{fig:basin_full}. The first heat-map sub-figure \ref{fig:basin_full}~(a) corresponds to $N_{bs}=8$, and it shows no variation in the color, indicating that all the 1000 samples lead to the non-chatter fixed point attractor. In other words, if the initial roughness of the workpiece is low or moderate, then the probability of noisy fluctuations pushing the dynamics to the vicinity of the limit cycle attractor is nearly zero. 

The heat-maps in Figs.~\ref{fig:basin_full}~(b)-(i) reveal that the number of samples transitioning to chatter vibrations $M_{CH}$ begins to rise as the number of basis functions $N_{bs}$ increases. This rise reflects how enhanced initial surface irregularities (represented by the history function) can affect the stability of the tool's motion, increasing its susceptibility to noise. However, when the initial tool vibration amplitude constraint $\beta$ is below 4, none of the simulations converged to the chatter limit cycle attractor, regardless of the specific $\alpha$ and $N_{bs}$ values. \textcolor{black}{In other words, the cutting process is stable when the amplitude of initial tool vibration is less than $2~mm$ (since $H_D=0.0005~m$, value of $\beta=4\times H_D = 2~mm$).} This indicates that one can suppress the noise-induced transition, especially in regions exhibiting multi-stability.  Moreover, as ${N_{bs}}$ increases, there is a consistent upward trend in the number of samples that end up in the sticking limit cycle attractor. This highlights the importance of surface complexity in pushing the system towards the chatter limit cycle. The other notable observation from Fig.~\ref{fig:basin_full}~(i) is the abrupt rise in $M_{CH}$ at $\beta=5$, jumping from zero to a maximum value of 20. This sudden change underscores the system's sensitivity to initial tool vibrations and the need for extra care in tool-workpiece interaction, particularly under stochastic conditions.

\subsection{Entropy Measures}
\begin{figure*}[!hbt]
     \centering
          \includegraphics[width=\textwidth]{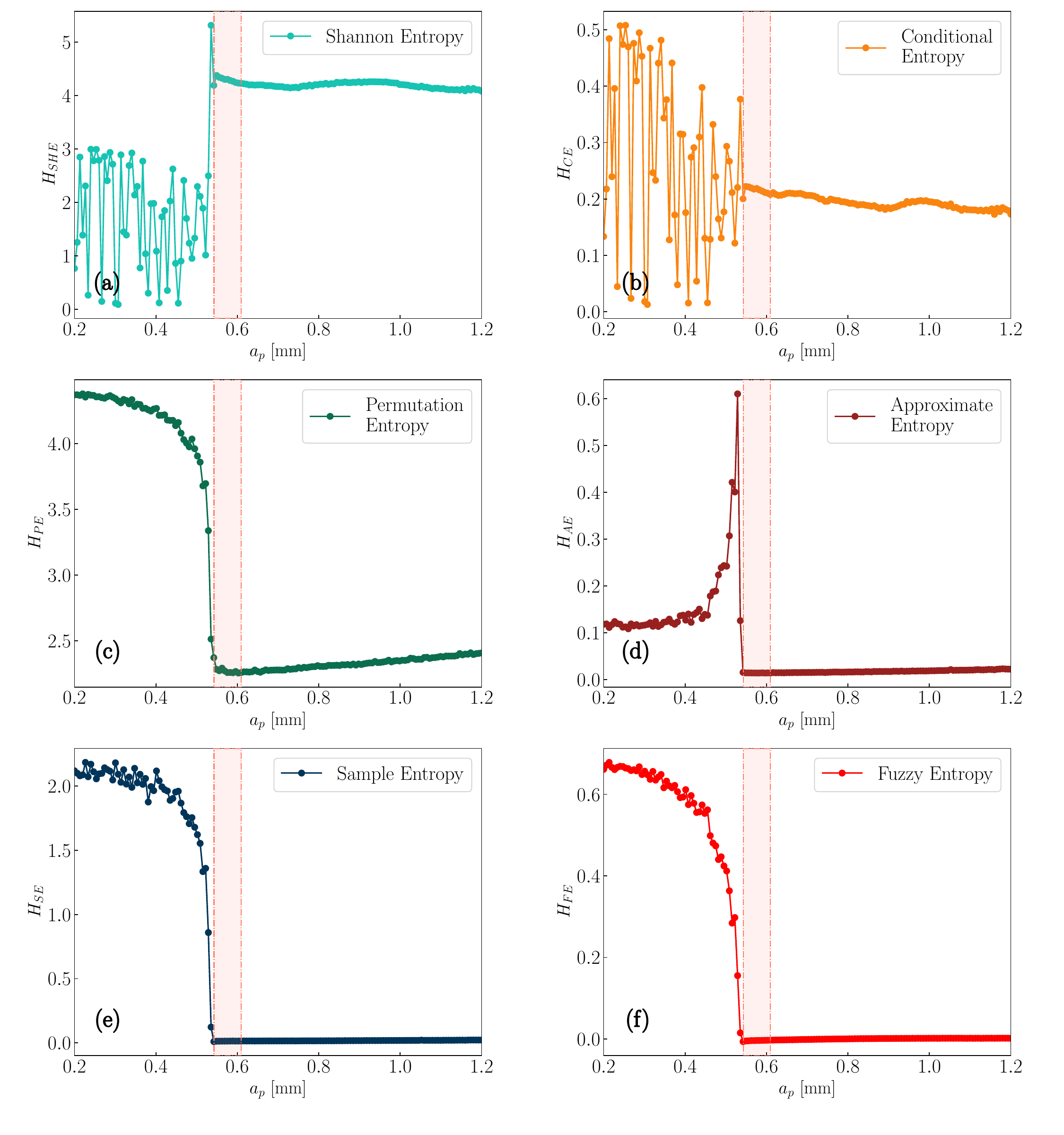}
         \caption{Various entropy measures. (a)~Shannon entropy, (b)~Conditional entropy, (c)~Permutation entropy, (d)~Approximate entropy, (e)~Sample entropy, and (f)~Fuzzy entropy. The shaded portion in all figures indicates the multi-stable zone from the deterministic study.}
         \label{fig:entropy}
\end{figure*}

\noindent Basin stability analysis provides the long-term stability structure of the system for a given initial workpiece profile. However, it is also important to compute the entropy measures to analyze the complexity and regularity of the noisy fluctuations in non-chatter, chatter, and multi-stable regimes. This section introduces and compares the result from several widely used entropy measures, namely, (a)~Shannon entropy, (b)~Conditional entropy, (c)~Permutation entropy, (d)~Approximate entropy, (e)~Sample entropy, and (f)~Fuzzy entropy,  each with distinct advantages in capturing features of given noisy data. All the entropy measures were computed using the comprehensive set of MATLAB routines provided in Ref.~\cite{monge2025entropy}. Fig.~\ref{fig:entropy} shows the results of entropy measures. Fig.~\ref{fig:entropy}~(a) represents the values of Shannon entropy ($H_{SHE}$) over the depth of cut values ($a_p$). Shannon entropy measures how much information is contained in the given signal \cite{shannon1948mathematical, porta1998measuring}. From Fig.~\ref{fig:entropy}~(a), it is clear that the $H_{SHE}$ fluctuates between 0 and 3 in the chatter-free region and settles around 4 in the chatter region. Also, it reaches a peak near  $a_p=0.54$ $mm$ at the onset of the multi-stable region (shaded portion in the figure). This change in the $H_{SHE}$ values is likely due to the presence of coexisting attractors, which increases the signal's unpredictability and complexity. The elevated entropy reflects the dynamical competition between the two attractors at the multi-stable zone. Fig.~\ref{fig:entropy}~(b) presents the results of conditional entropy. It quantifies the remaining uncertainty about the future state of a system given knowledge of its past states \cite{porta1998measuring}. It is clear from Fig.~\ref{fig:entropy}~(b) that the conditional entropy ($H_{CE}$) has values ranging from 0.01 to 0.5 in the non-chatter region and 0.2 to 0.22 in the multi-stable region.  A low $H_{CE}$ value indicates that the future state of the system is more predictable when the past states are known. Although only a fixed point attractor has been observed in the chatter-free region, noise alters the dynamics, causing the entropy values to oscillate. In the multi-stable and chatter region, the value is around 0.2 due to the system's periodic behavior. 
\par Fig.~\ref{fig:entropy}~(c) illustrates the results of permutation entropy ($H_{PE}$) measurement. Permutation entropy quantifies the complexity of a given signal or disorderliness based on its values' ordinal patterns \cite{bandt2002permutation}. A high value of $H_{PE}$ indicates high randomness of the signal. The permutation values in the chatter-free region are high when compared to the chatter and multi-stable regions. This is because added noise influences the patterns of the fixed point attractor in the chatter-free regions. Also, from Fig.~\ref{fig:entropy}~(c), it is observed that the $H_{PE}$ values reach the minima at the beginning of the multi-stable region, indicating the occurrence of high amplitude transitions between the underlying stable states. 
\par Fig.~\ref{fig:entropy}~(d) shows the values of approximate entropy $(H_{AE})$ computed varying the depth of cut value. Approximate entropy quantifies the amount of regularity in the given data \cite{pincus1991regularity}. This metric increases as $a_p$ increases in the stochastic fixed point regime. In the chatter-free region, $H_{AE}$ is around 0.1. However, it has the maximum value at the end of the chatter-free region, indicating that the system's predictability is very low near the multi-stable region. The reason for this is the co-existence of the two attractors, which affects the regularity of the time histories for the corresponding $a_p$ values. Further, the $H_{AE}$ values started to decrease and settled around $0.02$ after the chatter-free region, indicating that even though only high amplitude chatter vibrations were observed for $a_p$ from 0.61 to 1.2 $mm$, the dynamics of the system are not much affected by the included noise for that region. Thus, one observes fewer transitions from the stochastic limit cycle to the fixed point regime.
\par Fig.~\ref{fig:entropy}~(e) shows the results of sample entropy ($H_{SE}$) calculations. Sample entropy was introduced as an improvement over approximate entropy. It eliminates self-matching bias and has better relative consistency. Thus, a lower value of $H_{SE}$ indicates more self-similarity in the time series \cite{richman2000physiological}. It is clear from Fig.~\ref{fig:entropy}~(e) that $H_{SE}$ has the maximum value of 2 in the chatter-free region and starts to decrease until around $a_p=0.5$ $mm$, where the multi-stable region starts. Unlike approximate entropy, sample entropy shows no sudden jump to a higher value before entering the multi-stable region. This is due to the elimination of self-matching bias. It has the same values as the approximate entropy results for the chatter region. Fig.~\ref{fig:entropy}~(f) represents the fuzzy entropy calculations. Fuzzy entropy ($H_{FE}$) uses the concept of fuzzy sets, utilizing an exponential function to assess the vector similarity more smoothly instead of using a hard threshold \cite{chen2007characterization}. In this case, a trend similar to that of sample entropy has been observed, as sketched in Fig.~\ref{fig:entropy}~(f). 
\par It is also worth noting that Figs.~\ref{fig:entropy}~(a)~\&~(b), of Shannon entropy and the approximate entropy give contradictory results, {\it i.e.}, the value of $H_{SHE}$ is higher in chatter region than in the chatter-free region whereas in the approximate entropy calculations, chatter-free region has the higher value of $H_{AE}$ than in the chatter region. This is due to the nature of the entropy itself. When $H_{AE}$ is low and $H_{SHE}$ is high, it means that local patterns are regular, while the signal is globally diverse and random. The same behavior is observed when the results are compared between distribution-based ($H_{SHE}$ and $H_{CE}$) and pattern-based ($H_{AE}$, $H_{SE}$, and $H_{FE}$) entropy measurements which indicate that the introduced noise has more effect in the non-chatter region than in chatter region. 
\section{Conclusion}\label{conclusion}

\par This study proposes a model of an orthogonal metal cutting process combining regenerative, nonsmooth frictional, and stochastic effects. The role of nonsmooth friction in modelling the cutting system and predicting its stability is demonstrated through stability lobe diagrams obtained from a numerical continuation scheme. The proposed stochastic model demonstrates noise-induced transitions between stable fixed-point cutting and sticking limit-cycle oscillations associated with chatter vibrations. \textcolor{black}{The erosion of the stability region, when stochasticity due to correlated noise is present, is obtained and compared with the deterministic stability boundary.} Stochastic\textit{ P-bifurcations} as well as \textit{D-bifurcations }are observed to occur in the vicinity of the bi-stable region in the parameter space when the depth of cut is varied. The occurrence of \textit{P-bifurcations} is numerically quantified by a sudden expansion of the area under the \textit{j-pdf} at a designated level of probability, as the bifurcation parameter is varied. This area is numerically computed using the kernel density estimates. \textcolor{black}{The influence of varying time-delay in the stochastic machining model is studied as well.} Lyapunov exponents are computed in the weak sense by evaluating the growth rates of perturbations using the first statistical moment, specifically the ensemble-averaged expectations. The results of the stochastic bifurcation study revealed that the \textit{P-bifurcations} and \textit{D-bifurcations }occur at two distinct values of the parameter space. \textit{P-bifurcation} occurred when the depth of cut was 0.54 $mm$, whereas the\textit{ D-bifurcation} occurred when the depth of cut was 0.53 $mm$, despite being computationally computed at the same step size of 0.001. \textcolor{black}{A similar discrepancy within the values of $\pm 15 \, RPM$ was also observed when the spindle speed is varied.} This indicates that even though, in the averaged sense, the Lyapunov exponents can be negative, there could be a non-zero probability with which the orbit instantaneously transits to the other attractor.

\par Additionally, stability analysis of the stochastic attractors is carried out via two approaches, namely, basin stability and entropy measures. The basin stability analysis was carried out by considering history functions as sums of multiple harmonics generated via Fourier series expansions to incorporate the waviness of the initial workpiece surface and tool vibrations. The results of this analysis revealed that controlling chatter vibration is strongly linked to the magnitude of the initial tool vibrations. When the initial tool displacement constraint parameter was set to $\beta<4$, the system maintained stable cutting regardless of the other constraint parameter, $\alpha$, which is related to the workpiece surface. For values $\beta\ge4$, the chance of entering the chatter zone increased by $25\%$ to $100\%$ depending on the parameter value $\alpha$ and the number of basis functions of the Fourier series. Various entropy measures, namely, Shannon entropy, conditional entropy, permutation entropy, approximate entropy, sample entropy, and fuzzy entropy, were systematically evaluated across varying depths of cut values from 0.2 to 1.2 $mm$. Each measure captured distinct facets of system complexity. It was observed that entropy values tend to be higher in the noisy fixed-point regime than in the noisy limit-cycle regime, highlighting an inherent periodicity within the chatter-dominated behavior. 

\section*{Acknowledgments}
The authors acknowledge the support from Sonic Research IT HPC Cluster from UCD, funding from UCD Chang'an Dublin International College of Transportation, Research Ireland $22/FFP/P11457$ HarMonI, and Research Ireland funded I-Form Advanced Manufacturing Research Centre $1/RC/10295\_P2$.

\section*{Data Availability Statement}
The data that supports the findings of this study are available within the article.



\appendix
\section{Stability lobes calculation}\label{app1}
From Eq.~\eqref{eq-nondim2},
\begin{subequations}
    \begin{equation}\label{eq-ap-main1}
        \dot{y_1}(\tau) = y_2(\tau),
    \end{equation}
    \begin{equation}\label{eq-ap-main2}
        \dot{y_2}(\tau) = -y_1(\tau) - \xi y_2(\tau) + W [\mu_{vs} \cos(\gamma) - \sin(\gamma)]h(\tau) - Wc_y \dfrac{y_2(\tau)}{n},
    \end{equation}
\end{subequations}
where 
\begin{equation}
        \mu_{vs} = \mathrm{sign}(g)\left(\mu_d+(\mu_s-\mu_d)e^{|-g|}\right),
\end{equation}
\begin{equation}
  g = \frac{n}{v_s(\gamma,\phi)}-\nu \cos (\gamma) y_2(\tau),
\end{equation}
\begin{equation}
     \text{and} \, \,\,\,  h(\tau)=1-y(\tau)+y(\tau-\tau_w).
\end{equation}
For steady state,
\begin{equation}
    \left(y_1(\tau), y_2(\tau)\right) \equiv \left(y_{10}, 0\right),
\end{equation}
At the equilibrium chip thickness will be 
\begin{subequations}
    \begin{equation}
        h_0 = 1-y_{10} + y_{10} = 1,
    \end{equation}
    \begin{equation}
        g = \frac{n}{v_s}
    \end{equation}
     and the friction coefficient
    \begin{equation}\label{eq-ap-mu}
        \mu_{vs0} = \mu_d+(\mu_s-\mu_d)\exp{\left(-\dfrac{n}{v_s}\right)}.
    \end{equation}
\end{subequations}
Setting $\dot{y}_1=\dot{y}_2 = 0$, from Eq.~\eqref{eq-ap-main1},
\begin{equation*}
    \dot{y}_1 = y_2 = 0
\end{equation*}
and from Eq.~\eqref{eq-ap-main2},
\begin{equation*}
    y_{10} = W[\mu_{vs0}\cos(\gamma)-\sin(\gamma)]
\end{equation*}
substituting $\mu_{vs0}$ from Eq.~\eqref{eq-ap-mu} which gives
\begin{equation}\label{eq-ap-y10}
    y_{10} = W\left[\mu_d+(\mu_s-\mu_d)\exp{\left(-\dfrac{n}{v_s}\right)}\right]\cos(\gamma)-W\sin(\gamma).
\end{equation}
The next step is to linearize the system around the equilibrium. Let the system be perturbed slightly from equilibrium:
\begin{equation}
    y_1(\tau) = y_{10} + \delta y_1(\tau), \qquad y_2(\tau) = 0 + \delta y_2(\tau),
\end{equation}
where $\delta y1$ and $\delta y_2$ are small perturbations.
From Eq.~\eqref{eq-ap-main1}
\begin{equation}
    \delta \dot{y}_1 = \delta y_2,
\end{equation}
and from Eq.~\eqref{eq-ap-main2}
\begin{equation}\label{eq-ap-pert2}
    \delta \dot{y}_2 = - (y_{10}+\delta y_1) - \xi \delta y_2 + W(\mu_{vs} \cos(\gamma)-\sin(\gamma))h(\tau) - Wc_y \dfrac{\delta y_2}{n}.
\end{equation}
Substituting perturbations in chip thickness gives 
\begin{equation}
    h(\tau) = 1-\delta y_1 + \delta y_1(\tau - \tau_w).
\end{equation}
The friction coefficient depends on the velocity, so using the Taylor series to linearize it around $y_2=0$ gives
\begin{equation}\label{eq-ap-muvs}
    \mu_{vs} \approx \mu_{vs0} + \left.\dfrac{\partial \mu_{vs}}{\partial y_2}\right|_{y_2=0} \delta y_2 + \mathcal{O}(\delta y_2^2),
\end{equation}
where 
\begin{equation}\label{eq-ap-par-mu_vs}
    \dfrac{\partial \mu_{vs}}{\partial y_2} = \dfrac{\partial \mu_{vs}}{\partial g}\times\dfrac{\partial g}{\partial y_2}
\end{equation}

\begin{equation*}
    \begin{aligned}[b]
    \dfrac{\partial \mu_{vs}}{\partial g} = &\; \dfrac{\partial}{\partial g}\left[\mathrm{sign}(g)(\mu_d+(\mu_s-\mu_d)\exp(-|g|))\right]\\
    =&\; \dfrac{\partial}{\partial g}\left[\mathrm{sign}(g)(\mu_s-\mu_d)\exp(-|g|)\right]\\
    \end{aligned}
\end{equation*}
\begin{subequations}\label{eq-ap-parg}
    Case 1: $g>0 \rightarrow \mathrm{sign}(g)=+ve\; \mathrm{and}\;|g|=g $
\begin{equation}
    \begin{aligned}[b]
    \dfrac{\partial \mu_{vs}}{\partial g} 
    =&\; \dfrac{\partial}{\partial g}\left[+(\mu_s-\mu_d)\exp(-g)\right],\\
    =&\; -(\mu_s-\mu_d)\exp(-g),\\
    =&\; -(\mu_s-\mu_d) \exp(-|g|).
    \end{aligned}
\end{equation}
Case 2: $g<0 \rightarrow \mathrm{sign}(g)=-ve \; \mathrm{and}\;|g|=-g $
\begin{equation}
    \begin{aligned}[b]
    \dfrac{\partial \mu_{vs}}{\partial g} 
    =&\; \dfrac{\partial}{\partial g}\left[-(\mu_s-\mu_d)\exp(g)\right],\\
    =&\; -(\mu_s-\mu_d)\exp(g), \\
    =&\; -(\mu_s-\mu_d) \exp(-|g|).
    \end{aligned}
\end{equation}
\end{subequations}

\begin{equation}\label{eq-ap-pary2}
    \dfrac{\partial g}{\partial y_2} = - \nu \cos(\gamma).
\end{equation}
Substituting Eqs.~\eqref{eq-ap-parg} and \eqref{eq-ap-pary2} into Eq.~\eqref{eq-ap-par-mu_vs}, yields the partial derivative of the form
\begin{equation}
\begin{aligned}[b]
    \left.\dfrac{\partial \mu_{vs}}{\partial y_2}\right|_{y_2=0} =&\; (\mu_s-\mu_d) \nu \cos(\gamma) \exp{(-|[g]_{y_2=0}|)} \delta y_2, \\
    \left.\dfrac{\partial \mu_{vs}}{\partial y_2}\right|_{y_2=0} =&\; (\mu_s-\mu_d) \nu \cos(\gamma) \exp{\left(-\dfrac{n}{v_s}\right)} \delta y_2.
\end{aligned}\label{eq-ap-par-mu_vs_y20}
\end{equation}
Inserting the resulting partial derivative from Eq.~\eqref{eq-ap-par-mu_vs_y20} and $\mu_{vs0}$ from Eq.~\eqref{eq-ap-mu} into Eq.~\eqref{eq-ap-muvs}, yields the linearized friction coefficient of the form

\begin{equation}\label{eq-ap-muvs2}
    \mu_{vs} \approx \mu_d+ (\mu_s-\mu_d) \exp{\left(-\dfrac{n}{v_s}\right)} (1+ \nu \cos(\gamma)\delta y_2) 
\end{equation}
Substituting Eqs.~\eqref{eq-ap-y10} and \eqref{eq-ap-muvs2} into Eq.~\eqref{eq-ap-pert2}, followed by simplification and omission of second‑order terms, yields an equation of the form

\begin{subequations}
    \begin{equation}
        \delta \dot{y}_2 = -(1+Wa) \delta y_1(\tau) - (\xi + Wb)\delta y_2(t) + W a \delta y_1(\tau-\tau_w),
    \end{equation}
    For convenience, redefining $\delta y$
 as $ = y$ 
    \begin{equation}
         \dot{y}_2 = -(1+Wa) y_1(\tau) - (\xi + Wb) y_2(t) + W a  y_1(\tau-\tau_w),
    \end{equation}
\end{subequations}
where
\begin{subequations}
    \begin{equation}
        a = \left(\mu_d+(\mu_s-\mu_d)\exp{\left(-\dfrac{n}{v_s}\right)}\right)\cos(\gamma)-\sin(\gamma),
    \end{equation}
    \begin{equation}
        b = \dfrac{c_y}{n} + (\mu_d-\mu_s)\gamma \cos^2(\gamma)\exp{\left(-\dfrac{n}{v_s}\right)}.
    \end{equation}
\end{subequations}
Rearranging in matrix form
\begin{equation}
    \boldsymbol{\dot{y}}(\tau) = \boldsymbol{A}\boldsymbol{y}(\tau) + \boldsymbol{D}\boldsymbol{y}(\tau-\tau_w), 
\end{equation}
where 
\begin{equation}
    \boldsymbol{y} = \begin{pmatrix}
y_1 \\
y_2 
\end{pmatrix}, 
\boldsymbol{A} = \begin{pmatrix}
0 & 1 \\
-(1+Wa) & -(\xi+Wb)
\end{pmatrix}, 
\boldsymbol{D} = \begin{pmatrix}
0 & 0 \\
Wa & 0
\end{pmatrix}\label{eq-ap-matrices}
\end{equation}
The stability is then determined by the eigenvalues of the following characteristic equation.
\begin{equation}
    \det \left( \lambda\boldsymbol{I}- \boldsymbol{A}-\boldsymbol{D}e^{-\lambda \tau_w}\right) = 0, \label{eq-ap-char}
\end{equation}
Substituting $\lambda = \pm i\omega$ in Eq.~\eqref{eq-ap-char} and separating the real and imaginary parts gives the equations of the following form

\begin{subequations}\label{eq-ap-sldeqn}
    \begin{equation}
        \omega^2 - 1 - Wa[1-\cos(\omega\tau_w)] = 0,
    \end{equation}
    \begin{equation}
        \omega(\xi+W b) + Wa\sin(\omega \tau_w) = 0.
    \end{equation}
\end{subequations}

\section{Numerical continuation scheme} \label{app2}
Eq.~\eqref{eq-ap-sldeqn} is transcendental without an analytical solution. It is solved numerically using the scheme given in \cite{turning-yan2019modelling}. A detailed description of how the critical equations, Eq.~\eqref{eq-ap-sldeqn}, are solved is provided below. It begins by performing a systematic seed‑finding sweep over spindle speed and frequency ranges for a fixed value of depth of cut. Each valid intersection with a fixed depth-of-cut value serves as a starting point for continuation.

From these seeds, the algorithm traces entire stability boundaries using an arc‑length predictor–corrector method, ensuring smooth traversal of folds and turning points. The resulting curves are then converted to a dimensional form, sorted, filtered, and intersected to extract the upper envelope. This final envelope forms the complete stability lobe diagram. It is worth noting that this continuation method performs effectively for equations without nonsmoothness, as it smoothly follows the solution branch by appropriately adjusting the relaxation parameter.
\begin{algorithmic}[1]
\REQUIRE \textbf{Physical parameters:}\\
$\{m,c,\xi,k,K,c_y,R,H_D,\gamma,\phi,v_s,\nu, \mu_d,\mu_s\}$ \hfill$\triangleright$ Model data
\REQUIRE \textbf{Continuation parameters:}\\
inc\hfill$\triangleright$ Initial Predictor Step\\
rat \hfill$\triangleright$ Continuation Ratio\\
maxC \hfill$\triangleright$Max points per curve \\
maxT\hfill$\triangleright$ Second-point attempts \\
$\|F\|_{\text{tol}}$\hfill$\triangleright$ Residual tolerance\\
$\text{dup}_{\text{tol}}$ \hfill$\triangleright$ Duplicate tolerance\\
\REQUIRE $\bar{n}=[n_{\min},n_{\max},\Delta n]$, $\bar{\omega}=[\omega_{\min},\omega_{\max},\Delta\omega]$ \hfill$\triangleright$ Search grid
\REQUIRE $W^*$ \hfill$\triangleright$ Fixed value of $W$
\vspace{0.6em}

\COMMENT{\textbf{Phase 1:} \textit{Generate raw solution curves via arc-length continuation}}
\vspace{0.6em}
\STATE \texttt{CriticalEq} $(\omega,W,n)$: \hfill$\triangleright$ Function: Critical equations
\STATE \quad $a(n) = (\mu_d+(\mu_s-\mu_d)e^{-n/v_s})\cos\gamma - \sin\gamma$
\STATE \quad $b(n) = c_y/n + (\mu_d-\mu_s)\nu\cos^2\gamma\cdot e^{-n/v_s}$ \hfill$\triangleright$  $a$ and $b$ coefficients
\STATE \quad $\tau_w = 60/n$
\STATE \quad $F_1(\omega,W,n) = \omega^2-1-W\cdot a(n)\cdot(1-\cos(\omega\tau_w))$
\STATE \quad $F_2(\omega,W,n) = \omega(\xi+W\cdot b(n))+W\cdot a(n)\cdot\sin(\omega\tau_w)$

\STATE \RETURN $[F_1, F_2]$

\vspace{0.6em}
\STATE \texttt{Sol\_System}(F, guess, var, fixed):
\hfill$\triangleright$ Function: Nonlinear solver

\STATE $\mathbf{x} \gets$ vector from \texttt{guess[var]}
\hfill$\triangleright$ Initial iterate

\REPEAT
    \STATE Build dictionary $sol$ from $(var, \mathbf{x})$
    \hfill$\triangleright$ Insert free variables

    \IF{fixed $\neq$ \texttt{None}}
        \STATE $sol \gets sol \cup$ fixed
        \hfill$\triangleright$ Inject fixed parameters
    \ENDIF

    \STATE $R \gets F(sol)$
    \hfill$\triangleright$ Evaluate residual

    \IF{$\|R\| < \|F\|_{\mathrm{tol}}$}
        \STATE \textbf{return} $sol$
        \hfill$\triangleright$ Converged
    \ENDIF

    \STATE $J \gets$ finite-difference Jacobian of $F$
    \hfill$\triangleright$ Numerical derivative

    \STATE Solve $J\,\Delta x = -R$
    \hfill$\triangleright$ Newton step

    \STATE $x \gets x + \Delta x$
    \hfill$\triangleright$ Update iterate

\UNTIL max iterations reached

\STATE \textbf{return NULL}
\hfill$\triangleright$ No convergence (reject)

\vspace{0.6em}

\STATE \texttt{Calc\_Seed}$(F, W^*, \bar{n}, \bar{\omega})$:
\hfill$\triangleright$ Function: Seed finding

\STATE $\texttt{S} \gets \emptyset$
\hfill$\triangleright$ Initialize list

\FOR{$n_0 \in [n_{\min}, n_{\max}]$ (nstep)}
    \FOR{$\omega_0 \in [\omega_{\min}, \omega_{\max}]$ ($\omega$step)}

        \STATE $\texttt{guess} \gets \{n:n_0,\ \omega:\omega_0,\ W:W^*\}$
        \hfill$\triangleright$ Build initial guess

        \STATE $\texttt{sol} \gets \texttt{Sol\_System}(F,\texttt{guess},[n,\omega],\{W:W^*\})$
        \hfill$\triangleright$ Newton solve

        \IF{$\texttt{sol} \neq \text{NULL}$ \AND $n,\omega \in \mathbb{R}$ \AND $n \in [n_{\min},n_{\max}]$ \AND $\omega>0$}
            \STATE $\texttt{S}.\text{append}(\texttt{sol})$
            \hfill$\triangleright$ Accept valid seed
        \ENDIF

    \ENDFOR
\ENDFOR

\STATE Remove duplicates: $\|\mathbf{s}_i - \mathbf{s}_j\| < 10^{-12}, \mathbf{s} \in \texttt{S}$
\hfill$\triangleright$ Clean seed set

\STATE \RETURN $\texttt{S}$
\hfill$\triangleright$ All $(n,\omega)$ solutions at $W^*$

\vspace{0.6em}
\STATE \texttt{ArcConti}$(F, W^*, \bar{n}, \bar{\omega}, $\texttt{S}$, inc, rat, maxC, maxT)$:
\hfill$\triangleright$ Function: Arc-length continuation

\STATE \texttt{S\_Curves} $\gets \emptyset$, \quad varAll $=[n,\omega,W]$
\hfill$\triangleright$ Init

\FOR{$\texttt{\bf{S}}_0$ $\in$ $\texttt{S}$}

    \STATE sols $\gets$ $\texttt{S}_0$, \quad ratio $\gets$ rat
    \hfill$\triangleright$ Start new curve

    \STATE \textbf{Find 2nd point}
    \FOR{$k = 0$ to maxT$-1$}

        \STATE $\delta W_k \gets$ inc$/2^k$
        \hfill$\triangleright$ Shrinking step

        \STATE guess2$[W] \gets W^* + \delta W_k$
        \STATE sol2 $\gets$ \texttt{Sol\_System}$(F,$ guess2$, [n,\omega], \{W:W^*+\delta W_k\})$
        \hfill$\triangleright$ Newton solve

        \IF{$\|F(\texttt{sol2})\| < 10^{-10}$}
            \STATE sols.append(sol2)
            \STATE \textbf{break}
            \hfill$\triangleright$ Accept 2nd point
        \ENDIF

    \ENDFOR

    \STATE \textbf{Main loop}

    \FOR{step $=1$ to maxC}

        \STATE $s_1 \gets$ sols$[-2]$, \quad $s_2 \gets$ sols$[-1]$
        \STATE $\mathbf{v}_1=[s_1.n,s_1.\omega,s_1.W]$, \quad $\mathbf{v}_2=[s_2.n,s_2.\omega,s_2.W]$

        \STATE $\mathbf{v}_{g} \gets (1+\text{ratio})\mathbf{v}_2 - \text{ratio}\mathbf{v}_1$
        \hfill$\triangleright$ Predictor
            
        \STATE \texttt{eq\_aug}(sol):
        \hfill$\triangleright$ Augmented system for corrector
        
        \STATE $x \gets [\,sol[n],\ sol[\omega],\ sol[W]\,]$
        \hfill$\triangleright$ Current iterate
        
        \STATE $\text{ortho} \gets (x - v_g)\cdot(v_g - v_2)$
        \hfill$\triangleright$ Orthogonality constraint
        
        \STATE \textbf{return} $\big[F_1(sol),\ F_2(sol),\ \text{ortho}\big]$
        \hfill$\triangleright$ 3×3 augmented system
        \STATE sol $\gets$ \texttt{Sol\_System}$(eq\_aug, \mathbf{v}_g,$ varAll$)$
        \hfill$\triangleright$ Corrector

        \IF{sol = NULL}
            \STATE ratio $\gets$ ratio / rat
            \hfill$\triangleright$ Reduce step
            \STATE \textbf{continue}
        \ENDIF

        \STATE sols.append(sol)
        \STATE ratio $\gets$ rat
        \hfill$\triangleright$ Reset step size

        \IF{$n \notin [n_{\min}, n_{\max}]$}
            \STATE \textbf{break}
            \hfill$\triangleright$ Curve completed
        \ENDIF

    \ENDFOR

    \STATE \texttt{S\_Curves}.append(sols)
    \hfill$\triangleright$ Store completed curve

\ENDFOR

\STATE \RETURN \texttt{S\_Curves}
\hfill$\triangleright$ All continuation curves

\vspace{0.5cm}
\COMMENT{\textbf{Phase 2:} \textit{Lobe envelope construction}}
\vspace{0.6em}
\STATE \textbf{Dimensional conversion:}
\hfill$\triangleright$ Convert to physical units
\STATE $N \gets n\,\sqrt{k/m}$  \hfill$\triangleright$ Spindle speed (RPM)
\STATE $a_p \gets W\,\frac{k}{K}\,10^3$  \hfill$\triangleright$ Depth of cut (mm)

\STATE \textbf{Step 1:} 
\hfill$\triangleright$ Identify minima of each curve
\STATE $i_{\min} \gets \arg\min_j\, a_{p,i}(N_j)$ for all curves $\mathcal{C}_i$

\STATE \textbf{Step 2:} 
\hfill$\triangleright$ Sort lobes left→right
\STATE Sort curves by $N[i_{\min}]$ ascending

\STATE \textbf{Step 3:} 
\hfill$\triangleright$ Remove dominated lobes
\STATE Remove curves where $a_{p,i}(N_k) < a_{p,j}(N_k)+\epsilon$ for all common $N_k$

\STATE \textbf{Step 4:} 
\hfill$\triangleright$ Ensure increasing $N$
\STATE If a lobe has $N_0 > N_{\text{end}}$, reverse its arrays

\STATE \textbf{Step 5:} 
\hfill$\triangleright$ Compute lobe intersections
\STATE For each adjacent pair $(i,i+1)$ solve $a_{p,i}(N)=a_{p,i+1}(N)$

\STATE \textbf{Step 6:} 
\hfill$\triangleright$ Trim upper envelope
\STATE Cut each lobe at its intersection with the next

\STATE \RETURN SLD
\hfill$\triangleright$ Final stability lobe diagram

\end{algorithmic}

\bibliographystyle{elsarticle-num-names} 

\bibliography{references.bib}

\end{document}